\documentclass[12pt,sort&compress]{elsarticle}
\usepackage[utf8]{inputenc}

\usepackage{bm}
\usepackage{placeins}
\usepackage{booktabs} 
\usepackage{mathrsfs}
\usepackage{amsmath}
\usepackage{amsfonts}
\usepackage{amsthm}
\usepackage{upgreek}
\usepackage{color}
\usepackage{sansmath}

\usepackage{array}

\usepackage{floatrow} 
\usepackage{graphicx}
\usepackage[tableposition=top]{caption}

\usepackage{caption}
\usepackage{subcaption}
\usepackage{float}
\usepackage{fancyhdr}
\usepackage{url}
\usepackage{multicol}
\usepackage[top=1in,bottom=1in,left=1in,right=1in]{geometry}
\usepackage[pdfborder={0 0 0},colorlinks,allcolors=blue]{hyperref}
\usepackage{txfonts}
\usepackage{setspace}
\usepackage{arydshln}
\usepackage{enumitem}
\usepackage{lipsum}
\usepackage{siunitx,booktabs}
\usepackage{nth}
\usepackage{accents} 
\usepackage{multirow} 
\usepackage{float}  
\usepackage{floatrow} 
\theoremstyle{definition}

\allowdisplaybreaks

\usepackage{prettyref}
\newrefformat{fig}{Figure~\ref{#1}}
\newrefformat{Fig}{Fig.~\ref{#1}}
\newrefformat{Eq}{Eq.(\ref{#1})}
\newrefformat{eq}{Eq.(\ref{#1})}
\newrefformat{sec}{Section~\ref{#1}}
\newrefformat{tab}{Table~\ref{#1}}
\newrefformat{Tab}{Table~\ref{#1}}
\newrefformat{alg}{Algorithm~\ref{#1}}
\newrefformat{Alg}{Algorithm~\ref{#1}}

\usepackage[dvipsnames]{xcolor}
\newcommand\hl[1]{%
	\bgroup
	\hskip0pt\color{red!80!black}%
	#1%
	\egroup
}

\usepackage{algorithm}
\usepackage{algpseudocode}
\usepackage{algorithmicx}
\definecolor{shadecolor}{cmyk}{0,0,0,0.03}

\usepackage{amssymb}

 \usepackage{lineno}
\journal{Fluids}

\begin{document}
\begin{frontmatter}



\title{{\normalsize\textit{This article has been submitted to Fluids. Final version will appear at} \url{https://www.mdpi.com/journal/fluids}}\\[0.5em]
Computational study of vertical-axis MHK turbines using a coupled flow-sediment-turbine modeling approach}


\author[sbu]{Mehrshad Gholami Anjiraki}
\author[sbu]{Mustafa Meriç Aksen}
\author[sbu]{Samin Shapourmiandouab}
\author[sbu]{Jonathan Craig}
\author[sbu]{Ali Khosronejad\corref{cor1}}
\cortext[cor1]{Corresponding author: \href{mailto:ali.khosronejad@stonybrook.edu}{ali.khosronejad@stonybrook.edu}}

\address[sbu]{Department of Civil Engineering, Stony Brook University, Stony Brook, NY 11794, USA}

\begin{abstract}
We present a coupled large-eddy simulation (LES) and bed morphodynamics study to investigate the influence of sediment dynamics on the performance of a utility-scale marine hydrokinetic vertical-axis turbine (VAT) parametrized by an actuator surface model. By resolving the two-way interactions between turbine-induced flow structures and bed evolution, the study offers insights into the environmental implications of VAT deployment in riverine and marine settings. A range of tip speed ratios is examined to evaluate wake recovery, power production, and bed response. The actuator surface method (ASM) is implemented to capture the effects of rotating vertical blades on the flow, while the immersed boundary method accounts for fluid interactions with the channel walls and sediment layer.
The results show that higher TSRs intensify turbulence, accelerate wake recovery over rigid beds, and enhance erosion and deposition patterns beneath and downstream of the turbine under live-bed conditions. Bed deformation under live-bed conditions induces asymmetrical wake structures through jet flows, further accelerating wake recovery and decreasing turbine performance by about $2\%$, compared to rigid-bed conditions. Considering the computational cost of the ASM framework, which is nearly $4\%$ of the turbine-resolving approach, it provides an efficient yet robust tool for assessing flow-sediment-turbine interactions.
\end{abstract}


\end{frontmatter}

\section{Introduction}

In recent years, renewable energy has seen rapid growth and become a major focus for global investment~\cite{[1],[2]}. Hydrokinetic devices, in particular, are gaining attention because of their emerging potential to generate power from the natural flow of rivers, tides, and ocean currents. Unlike conventional hydropower systems that rely on dams or penstocks to create artificial water heads, hydrokinetic systems extract energy directly from moving water without disrupting natural waterways~\cite{[4],[5],[6],[7]}. Therefore, marine energy is a promising power source, offering high reliability, predictability, and consistent output~\cite{[8],[9],[10]}.
Despite its potential, tidal stream technology faces certain challenges, including high construction and maintenance costs due to its intricate and underexplored interactions with marine environments ~\cite{[11],[13],[14],[15]} and a lack of a well-established supply chain~\cite{[11],[12]}. Although tidal stream turbines are often considered viable only in select regions, demonstration projects in the US and Europe~\cite{[16],[17],[18],[19],[20],[21]} suggest that continued technological advances are unlocking the global potential of tidal stream turbines~\cite{[10],[13]}.
Although riverine energy is smaller in scale than ocean-based sources, its technical energy production is considerable: estimated at 99 TWh per year in the US alone, riverine energy may power approximately $9.3$ million homes~\cite{[8]}. Pilot deployments of marine turbines in riverine environments have shown that this relatively untapped resource is feasible and effective~\cite{[20],[80]}. Throughout marine energy research so far, horizontal-axis turbines (HATs) have dominated tidal energy research, comprising roughly $75\%$ of all studies in the field~\cite{[21],[22],[23]}. This focus is largely due to the successful transfer of design principles from wind turbines, which has accelerated convergence and optimization in marine HAT development~\cite{[23],[24]}.

At the same time, vertical-axis turbines (VATs) have gained attention in recent renewable energy research~\cite{[4]} due to several practical and environmental advantages in comparison to HATs ~\cite{[8],[13],[34],[4b],[27],[5b]}. 
While VATs are generally less efficient and not self-starting like HATs~\cite{[26]}, they offer operational benefits in array configurations, such as faster wake recovery~\cite{[27],[32]}, which can improve overall farm performance. Their design, featuring single-axis rotation, helps them manage unsteady fluid forces effectively~\cite{[25]}. Due to their larger swept area and high power density, VATs are well-suited for shallow, near-shore waters and can achieve greater power output in such conditions~\cite{[13],[28],[29]}. Their omnidirectional flow acceptance~\cite{[23],[24],[29]} and their relatively simple installation requirements~\cite{[30]} make them especially suitable for riverine environments. VATs also tend to have reduced ecological and hydrological impact~\cite{[23],[34]}, further supporting their use in sensitive or shallow aquatic sites.
Extensive experimental research and design optimization efforts have advanced VAT technology~\cite{[34],[35],[37],[25],[29],[38],[39],[40],[41],[42]}, and they have been successfully deployed in industrial applications~\cite{[36]}.

Since riverine currents are unidirectional, it is easier to model and manage them within experimental studies. Nevertheless, turbine installations in rivers can significantly impact river morphology. This poses challenges not only for system design and environmental impact mitigation, but also for modeling. Given the high cost and time demands of physical experiments, computational fluid dynamics (CFD) is often used to complement laboratory studies to explore a wider range of scenarios~\cite{[13],[27]}. 
Together with experimental models, computational studies on marine VATs are expanding to improve understanding of their performance and wake dynamics. Large-eddy simulation (LES) has emerged as a reliable model for CFD simulations, showing strong agreement with theoretical predictions~\cite{[44]} and outperforming unsteady Reynolds-averaged Navier–Stokes (URANS) in capturing unsteady flow behavior, as demonstrated in vertical-axis wind turbine (VAWT) studies~\cite{[45]}. For example, \citet{[46]} applied LES to a wind tunnel study to discover that at high Reynolds numbers, lower tip speed ratios (TSRs) produce more asymmetric VAWT wakes and larger vortices due to stronger dynamic stall. \citet{[49]} employed LES to examine the effect of wind speed on VAWT output under different pitch conditions. \citet{[27]} coupled LES with the Immersed Boundary Method (IBM) to simulate Darrieus-type VATs in turbulent flow, validating their results against body-fitted methods and experimental data across various TSRs. Further research by \citet{[50]} applied LES-IBM to observe that higher blockage ratios can enhance downstream turbine performance in turbine arrays. Based on LES-IBM simulations, ~\citet{[51]} found that higher dynamic solidity enhances wake recovery. 

For all CFD simulations of turbines, there are diverse methods to represent the turbine geometry according to modeling accuracy and computational resources. While turbine-resolving methods provide high-fidelity results with minimal modeling assumptions~\cite{[143]}, their high computational cost limits widespread use. To accommodate computational resource limits, recent studies on VATs have increasingly adopted actuator-based approaches to evaluate turbine performance, wake dynamics, and flow interactions in both atmospheric and hydrokinetic contexts.
For instance, \citet{[48]} coupled the actuator line method (ALM) with LES to study VAWT wakes within the atmospheric boundary layer (ABL), finding that coarse-resolution simulations in large wind farms fail to adequately capture individual turbine wake structures. Similarly, \citet{[47]} used LES-ALM to optimize turbine solidity and tip speed ratio (TSR) for maximum power output and to identify regions of highest velocity deficit and turbulence within the wake. \citet{[156]} developed a three-dimensional (3D) simulation framework that couples the ALM with Volume of Fluid (VOF) modeling to evaluate VAT array performance in open-channel flows. \citet{[158]} employed URANS simulations with an actuator disk method (ADM) to compare array configurations, finding that side-by-side layouts consistently outperformed staggered ones across a range of TSRs and flow directions.
\citet{[159]} validated LES-ALM simulations against experimental data to demonstrate accurate predictions of flow behavior and power output whilst reducing computational costs by two orders of magnitude compared to fully resolved blade models. Subsequently, \citet{[160]} employed LES–ALM to compare straight and helical blade designs, and thereby they observed that helical blades intensify near-wake velocity fluctuations and generate secondary vertical flow motion.
~\citet{[162]} reported relatively fast wake recovery in VATs according to experimental results with Reynolds-averaged Navier–Stokes (RANS)–ADM simulations despite the ADM’s limited resolution of detailed wake structures. ~\citet{[163]} supported this observation, concluding that actuator-based models, particularly ALM, closely approximate the outcomes of full CFD while remaining computationally feasible. 
~\citet{[167]} used RANS–ALM to support that dual-rotor counter-rotating VATs could boost individual turbines' performance. Similarly, \citet{[168]} employed ADM to evaluate blockage effects on turbine performance and found that as blockage intensifies, the extracted power scales at a greater than cubic rate with respect to the upstream velocity.

Notwithstanding the ALM's viability, several studies have also demonstrated the efficacy of the actuator surface method (ASM) for modeling VATs. Most extant ASM studies examine VAWTs instead of marine VATs; nevertheless, they present valuable implications and considerations for marine settings. An early work by ~\citet{[170]} integrated Patankar’s SIMPLER algorithm ~\cite{[171]} with ASM to investigate the wake flow and performance of a two-dimensional ($2D$) VAWT. This approach was later extended to $3D$ ASM for a curved-blade VAWT by ~\citet{[172]}. ~\citet{[173]} solved the $2D$ compressible laminar Euler equations using a finite difference method, coupled with the ASM, to model the flow field around a VAWT. ~\citet{[174]} applied a $2D$ ASM coupled with RANS to simulate the wake of a two-bladed VAWT with a NACA$0015$ airfoil, reporting good agreement with experimental data. Shamsoddin and Porté-Agel ~\cite{[165],[166]} combined LES with both ALM and ASM to simulate $3D$ wake structures, to validate blade forces via blade-element theory, and to assess optimal combinations of TSR and solidity. Their findings also highlighted ASM's lower grid-resolution requirements, which make it particularly favorable for wind farm modeling. Through LES-ASM, ~\citet{[23]} verified improved accuracy in capturing upstream blade blockage effects compared to ALM, although turbulent kinetic energy (TKE) was underestimated with coarse grids. Notably, they observed only marginal improvement with further mesh refinement, underscoring the ASM’s effectiveness on relatively coarse grids. More recently, ~\citet{[176]} employed the ASM with an explicit momentum solver to study VAT wake flows and validated their approach with experimental measurements. They further emphasized that AS addresses key shortcomings of momentum-based models, such as the double-multiple streamtube (DMST) model, which often fail to accurately capture complex wake dynamics.


Interactions between marine hydrokinetic (MHK) VATs and bed morphodynamics have become a key research challenge, with growing concern over their uncertain impacts in riverine and tidal settings ~\cite{[56]}. By altering sediment transport dynamics, marine hydrokinetic turbines may substantially alter seabed morphology with the potential of disrupting benthic habitats and water quality \cite{[57],[58],[59],[60],[61]}. Although early modeling efforts provided valuable insights into ecological and hydrodynamic changes, they often lacked field validation and failed to adequately address sediment transport~\cite{[62]}, primarily due to the greater emphasis on power optimization in early environmental assessments~\cite{[63]}. To address this discrepancy, analogies have emerged between MHK turbines and structures, including bridge piers and offshore turbines, which inform possible local scour and debris effects for MHK turbines~\cite{[64], [65],[66],[67],[68],[69]}. As noted by ~\citet{[71]}, turbine installation using anchors and moorings can disturb sediments, causing short-term turbidity and contaminant release, while long-term impacts arise from wake turbulence generated during rotor operation. Furthermore, sediment transport can become an obstacle in itself by increasing turbine fatigue and maintenance costs through turbine-sediment interactions ~\cite{[70]}. Previous studies on marine HATs have shown that they can intensify flow velocity and sediment erosion, particularly under conditions of reduced tip clearance ~\cite{[72]}. ~\citet{[73]} found that sediment transport lowers turbine performance, with larger rotors and steeper bedforms increasing sediment interaction. ~\citet{[74]}  experimentally indicated that reduced tip clearance in HATs leads to greater scour, while
~\citet{[68]} numerically indicated that altered morphodynamics and debris affect utility-scale HAT power output. ~\citet{[62]} used small-scale HAT models in lab experiments to link scour depth with turbine-induced drag, revealing both local and non-local effects on bed morphodynamics. Afterward, ~\citet{[76]} explored how asymmetrically placed HATs affect morphodynamics. Further studies ~\cite{[62],[72],[73]} showed that MHK HATs influence both local sediment transport and broader bedform migration. Although no damage to turbines was observed, ~\citet{[62]} warned that altered sediment patterns could threaten river morphology. 

Research on VAT–sediment interactions is relatively recent. ~\citet{[77]} showed that different VATs significantly impact local morphodynamics and downstream shear stress.~\citet{[78]} proposed a drag-driven VAT design for live-bed conditions. ~\citet{[66]} identified erosion-prone areas on VATs exposed to suspended sediments and suggested protective measures. Given the complex interactions between VATs and morphodynamics, several studies have been conducted to better understand these relatively underexplored effects. For instance, ~\citet{[79]} recommended mid-depth turbine deployment to reduce VAT-bed interaction. ~\citet{[155]} investigated the two-way interactions between VAT wake flow and sediment transport, examining both the impact of the wake on sediment dynamics and the influence of evolving live-bed conditions on VAT performance and wake structure. However, their use of a geometry-resolving method for turbine modeling makes the approach computationally expensive, limiting its practicality for studying different utility-scale VATs. Due to limited research, accurate and cost-effective modeling of the two-way interaction between VATs and sediment transport remains a major challenge. Further computational studies are essential to bridge this gap and support the practical development of tidal farms.

This study aims to explore the two-way interactions between a utility-scale VAT and sediment transport. We simulated live-bed sediment dynamics by considering a typical sand particle size commonly found in natural riverbeds~\cite{[169]}. It examines how different TSRs affect bed evolution (e.g., sand wave dynamics) and how evolving bed topography, in turn, influences turbine performance and wake flow. Simulations under both rigid- and live-beds enable a comparative analysis to better understand VATs’ environmental impacts. Numerical simulations were conducted using the in-house virtual flow simulator (VFS)-Geophysics code~\cite{[107], [147]}, which couples hydrodynamics with bed morphodynamics. The turbulence is captured using the LES method. A wall model is employed to reduce the computational cost of LES at high Reynolds numbers. The ASM is used to model turbine blades, and the immersed boundary method handles the evolving bed geometry~\cite{[101], [113]}. Sediment dynamics are captured by solving the sediment mass balance equation within the bedload layer using a dual time-stepping scheme ~\cite{[69], [128]}, and a sand slide model is implemented to maintain physically realistic bed slopes.

The paper is organized as follows: \prettyref{sec:2} outlines the governing equations, \prettyref{sec:3} details the test case setup and sediment transport model, \prettyref{sec:4} presents and discusses results, and \prettyref{sec:5} summarizes key findings and implications.

\section{Governing equations}
\label{sec:2}
\subsection{The hydrodynamic model}
\label{sec:2.1}
\noindent The hydrodynamics model solves the spatially filtered Navier-Stokes equations for incompressible flow in non-orthogonal generalized curvilinear coordinates. In compact Newton notation, with repeated indices indicating summation, the equations are expressed as follows ~\cite{[101], [135], [179], [180]}:

\begin{equation}
    J\frac{\partial U^j}{\partial\xi^j}=0
    \label{eq:1}
\end{equation}

\begin{equation}
    \frac{\partial U^i}{\partial t}=\frac{\xi_l^i}{J}\left(\frac{\partial}{\partial\xi^j}\left(U^ju_i\right)+\frac{1}{\rho}\frac{\partial}{\partial\xi^j}\left(\mu\frac{G^{jk}}{J}\frac{\partial u_i}{\partial\xi^k}\right)-\frac{1}{\rho}\frac{\partial}{\partial\xi^j}\left(\frac{\xi_i^jp}{J}\right)-\frac{1}{\rho}\frac{\partial\tau_{ij}}{\partial\xi^j} + F_{\mathrm{ext}}
\right)
    \label{eq:2}
\end{equation}

\noindent where the Jacobian of the geometric transformation, $J=\left|\partial\left(\xi^1,\xi^2,\xi^3\right)/\partial\left(x_1,x_2,x_3\right)\right|$, transforms the coordinate system from Cartesian into curvilinear. The contravariant volume flux is $U^i=(\xi_m^i/J)\ u_m,$, where $\xi_l^i=\partial\xi^i/\partial x_l$. The $i$-th filtered velocity component in Cartesian coordinates is denoted by $u_i$, and $\mu$ represents the dynamic viscosity of the fluid (i.e., water).  
The contravariant metric tensor is $G^{jk}=\xi_l^j\ \xi_l^k$.
The fluid density is $\rho = 1000 \ \text{kg/m}^3$, and $p$ represents the pressure field.  
The subgrid-scale stresses are modeled using the dynamic Smagorinsky formulation within the LES turbulence model, defined as \cite{[100], [103], [104]}:

\begin{equation}
     \tau_{ij}=-2\mu_{\mathrm{t}} \overline{S}_{{ij}}+\frac{1}{3}\tau_{kk}\delta_{ij} 
     \label{eq:3}
\end{equation}
\begin{equation}
    \mu_{\mathrm{t}} = C_{\mathrm{s}}\Delta^2 \left |\overline{S} \right|
    \label{eq:4}
\end{equation}

\noindent where $\mu_{\mathrm{t}}$ denotes the eddy viscosity, $\overline{S}_{ij}$ is the filtered strain-rate tensor, and $\delta_{ij}$ is the Kronecker delta. The Smagorinsky constant is represented by $C_{\mathrm{s}}$, and the strain-rate magnitude is
$\left| \overline{S} \right| = \sqrt{2 \, \overline{S}_{ij} \, \overline{S}_{ij}}.$
The filter width, $\Delta$, is taken as the cubic root of the cell volume, such that $\Delta = J^{-1/3}$ \cite{[100]}.

\subsection{Turbine modeling}
\label{sec:2.2}
\noindent Two widely adopted strategies are commonly used to simulate flow–turbine interactions ~\cite{[143]}. The first strategy involves turbine parameterization techniques such as actuator models, which introduce lift and drag forces as body force terms into the governing flow equations to represent the turbine's influence as momentum sinks (see \prettyref{eq:2}) ~\cite{[33], [137], [138], [180]}. The second strategy is the geometry-resolving approach ~\cite{[139], [140], [141], [142], [143], [155], [179]} which explicitly resolves the turbine blades’ geometry using sufficiently fine computational grids to capture fluid–structure interactions in detail. Although this latter approach offers high-fidelity results with minimal modeling assumptions ~\cite{[143]}, its significant computational cost limits its feasibility for turbine simulations. Therefore, in this study, we employ the ASM, which offers a balanced trade-off between computational efficiency and the resolution of turbine-induced flow features whilst maintaining acceptable accuracy even on coarse meshes ~\cite{[23]}.

\subsubsection{Actuator surface method}

\noindent The ASM represents turbine blades and nacelles as immersed surfaces that apply lift and drag forces to the flow. These forces are computed over a Lagrangian mesh embedded in the fluid domain and incorporated into the momentum equations, enabling efficient simulation of the effects of rotating blades on the flow field without resolving the full blade geometry. Using the blade element theory ~\cite{[178]}, the lift and drag forces are calculated as follows ~\cite{[177]}:

\begin{equation}
  F_{\mathrm{D}}=\frac{c}{2}\rho C_{\mathrm{D}}V_{\mathrm{rel}}^2
  \label{eq:6}
\end{equation}
\begin{equation}
   F_{\mathrm{L}}=\frac{c}{2}\rho C_{\mathrm{L}}V_{\mathrm{rel}}^2
   \label{eq:7}
\end{equation}

Here, $c$ denotes the turbine blade chord length while $C_{\mathrm{D}}$ and $C_{\mathrm{L}}$ correspond to the drag and lift coefficients which depend on the Reynolds number and the angle of attack ~\cite{[33]}. For an in-depth discussion on tip-loss effects and the corresponding corrections to the lift and drag coefficients, the reader is referred to Ref.~\cite{[33]}. The relative velocity, $V_{\mathrm{rel}}$, is defined as follows ~\cite{[177]}:

\begin{equation}
V_{\mathrm{rel}} = \left(u_x, u_\theta - \Omega r\right)
\label{eq:8}
\end{equation}

\noindent Here, $\Omega$ denotes the rotor’s angular velocity, and $r$ represents the radial distance from the rotor center to the blade element at the same vertical elevation, which corresponds to the turbine radius. The velocity components $u_x$ and $u_\theta$ represent the axial and azimuthal flow components, respectively.
These components are defined as follows ~\cite{[177]}:

\begin{equation}
u_x = \mathbf{\bm{u}}\left(X\right) \cdot \mathbf{\bm{e}}_x
\label{eq:9}
\end{equation}

\begin{equation}
u_\theta = \mathbf{\bm{u}}\left(X\right) \cdot \mathbf{\bm{e}}_\theta
\label{eq:10}
\end{equation}

Here, $\mathbf{e}_x$ and $\mathbf{e}_\theta$ denote the unit vectors in the axial and azimuthal directions, respectively. The variables $x_i$ and $X_i$ (with $i = 1, 2, 3$ in compact tensor notation) represent the Cartesian and Lagrangian coordinate systems, corresponding to the fluid grid nodes and the actuator surface points, respectively. The discrete delta function $\delta_h$ is used to transfer the velocity from the background grid onto a set of Lagrangian points distributed along a line located on the blade surface~\cite{[177],[149]}:

\begin{equation}
\mathbf{\bm{u}}(X) = \sum_{N_f} \mathbf{\bm{u}}(\mathbf{\bm{x}}) \delta_h(x - X) V(x)
\label{eq:11}
\end{equation}

\begin{equation}
\delta_h(x-X) = \frac{1}{V} \phi\left(\frac{x-X}{h_x}\right) \phi\left(\frac{y-Y}{h_y}\right) \phi\left(\frac{z-Z}{h_z}\right)
\label{eq:12}
\end{equation}

\noindent Here, $N_f$ denotes the total number of fluid cells, $V(x)$ represents the volume of a fluid cell at position $x$ ~\cite{[137]}, and the function $\phi$ corresponds to the smoothed four-point cosine kernel ~\cite{[55]}.
Assuming a uniform distribution of force along the chord length, the force per unit area can be expressed by the following equation ~\cite{[177]}:

\begin{equation}
f(X) = \frac{\left(F_{\mathrm{L}} + F_{\mathrm{D}}\right)}{c}
\label{eq:14}
\end{equation}

Once all parameters from \prettyref{eq:6} to \prettyref{eq:14} are evaluated, the distributed body force can be computed as follows ~\cite{[177]}:

\begin{equation}
f_{\mathrm{AS}}(x) = \sum_{N_{\mathrm{s}}} f(X) \delta_h(x-X) A(X)
\label{eq:15}
\end{equation}

\noindent where $A(X)$ denotes the area of the actuator surface grid cells and $N_{\mathrm{s}}$ represents the total cells on the actuator surface grid. The resulting distributed body force, $f_{\mathrm{AS}}$, is then introduced as a source term on the right-hand side of \prettyref{eq:2}, as the external forces per unit volume ($F_{\mathrm{ext}}$), representing the influence of the rotating blades on the surrounding fluid nodes. Finally, the ASM has been validated in multiple studies \cite{[33], [181], [182]}, and the grid sensitivity has been examined through dedicated analyses \cite{[183]}.








\subsection{Bed morphodynamics}
\noindent The non-cohesive bed material, consisting of sand with a $d50$ of $0.7mm$, was considered. Sediment transport occurs through rolling or sliding, saltation, or suspension, depending on the bed shear velocity relative to the critical threshold \cite{[155]}. As described by \citet{[105]}, sediment transport can be classified into three modes: bed load, which includes rolling and saltation; suspended load, where particles are maintained in motion by turbulence; and wash load, which consists of fine particles that remain continuously in transport. The temporal change in the bed elevation (i.e., the elevation of the sediment/water interface), $Z_{\mathrm{b}}$, is governed by the non-equilibrium Exner-Polya mass balance equation, as follows~\cite{[106]}:

\begin{equation}
(1-\gamma)\frac{\partial Z_{\mathrm{b}}}{\partial t} + \frac{1}{A_h} \nabla \cdot \mathbf{\bm{q}}_{\mathrm{BL}} = {D_b}-{E_b}
\label{eq:6_1}
\end{equation}

\noindent Here, $\gamma$ denotes the sediment porosity ($=0.4$). $D_b$ and $E_b$ represent the net rates of sediment deposition and entrainment, respectively~\cite{[107]}. In this study, both $D_b$ and $E_b$ are set to zero, since suspended load transport is not considered. The divergence operator $\nabla$ is applied to the bed load flux vector $\bm{q}_{\mathrm{BL}}$, representing net sediment transport within the bed load layer. The bed load flux vector is defined as follows~\cite{[108]}:

\begin{equation}
\mathbf{\bm{q}}_{\mathrm{BL}} = \psi \, {\|d_{\mathrm{s}}\|} \, {\|\delta_{\mathrm{BL}}\|} \, \mathbf{\bm{u}}_{\mathrm{BL}}
\label{eq:7_1}
\end{equation}

\noindent where $d_s$ denotes the edge length of each triangular bed mesh element, $\delta_{\mathrm{BL}} = 0.005$ is the thickness of the bed load layer, $\bm{u}_{\mathrm{BL}}$ represents the velocity component parallel to the bed surface, and $\psi$ is the sediment concentration defined as follows:

\begin{equation}
\psi = 0.015 \frac{d_{50}}{\delta_{\mathrm{BL}}} \frac{T^{3/2}}{D_*^{3/10}}
\label{eq:7_1}
\end{equation}

\begin{equation}
D_* = d_{50} \left[ \left( \frac{\rho_{\mathrm{s}} - \rho}{\rho v^2}g \right)^{1/3} \right]
\label{eq:9}
\end{equation}

\noindent where $d_{50}$ is the mean grain size, $\rho_{\mathrm{s}} = 2650\ kg/m^3$ is the sediment particle density, and $\nu$ denotes the kinematic viscosity of water. $T$ denotes he non-dimensional excess shear stress and is defined as follows:

\begin{equation}
T = \frac{\tau_* - \tau_{*\mathrm{cr}}}{\tau_{*\mathrm{cr}}}
\label{eq:10}
\end{equation}

\noindent where $\tau_*$ is the bed shear stress and $\tau_{*\mathrm{cr}}$ is the critical bed shear stress. More details are provided in Ref.~\cite{[101],[105],[107],[108],[109],[110],[111],[112]}.

The bed load sediment flux ($\mathbf{\bm{q}}_{\mathrm{BL}}$) is computed at cell faces using a second-order GAMMA differencing scheme \cite{[112]} after obtaining all parameters from cell centers. To ensure a physically realistic bed topography, a mass-balanced sand slide model is applied when bed local slopes exceed the angle of repose \cite{[107], [101], [112], [113], [135]}. The model redistributes excess sediment from flagged cells to neighbors iteratively, until all local slopes fall below $99\%$ of the angle of repose \cite{[155]}.

\subsection{The coupled hydro- and morpho-dynamics}

\noindent Two primary fluid structure interaction (FSI) coupling approaches are used to model hydro- and morpho-dynamic interactions: strong and loose coupling \cite{[108]}. Strong coupling updates boundary conditions iteratively within each time step (implicit in time), offering higher stability but at a greater computational cost. Loose coupling, by contrast, updates boundaries using solutions from the previous time step (explicit in time), avoiding extra iterations and improving efficiency. Given its computational advantages and demonstrated robustness in similar studies \cite{[155]}, the loose coupling method is adopted in this work \cite{[101], [107], [111], [112], [113], [135]}.

\noindent Due to the significant difference in convergence time scales between hydrodynamics (seconds to minutes) and morphodynamics (hours to days), a dual time-stepping method is employed~\cite{[128]}, allowing the use of larger time steps for the morphodynamic solver. In this study, the morphodynamic solver employs a time step that is two orders of magnitude greater than that of the flow solver \cite{[155]}.

\noindent The flow–morphodynamics coupling within the curvilinear immersed boundary (CURVIB) framework involves solving hydrodynamic and morphodynamic equations in separate domains, while incorporating boundary conditions from the other domain at the sediment-water interface. More specifically,  the flow solver implements updated bed elevations and bed vertical velocities as boundary conditions, whereas the morphodynamic solver uses flow velocities and shear stresses from the hydrodynamics solver~\cite{[155]}.

\section{Test case description and computational details}
\label{sec:3}

\begin{figure} [H]
  \includegraphics[width=1\textwidth]{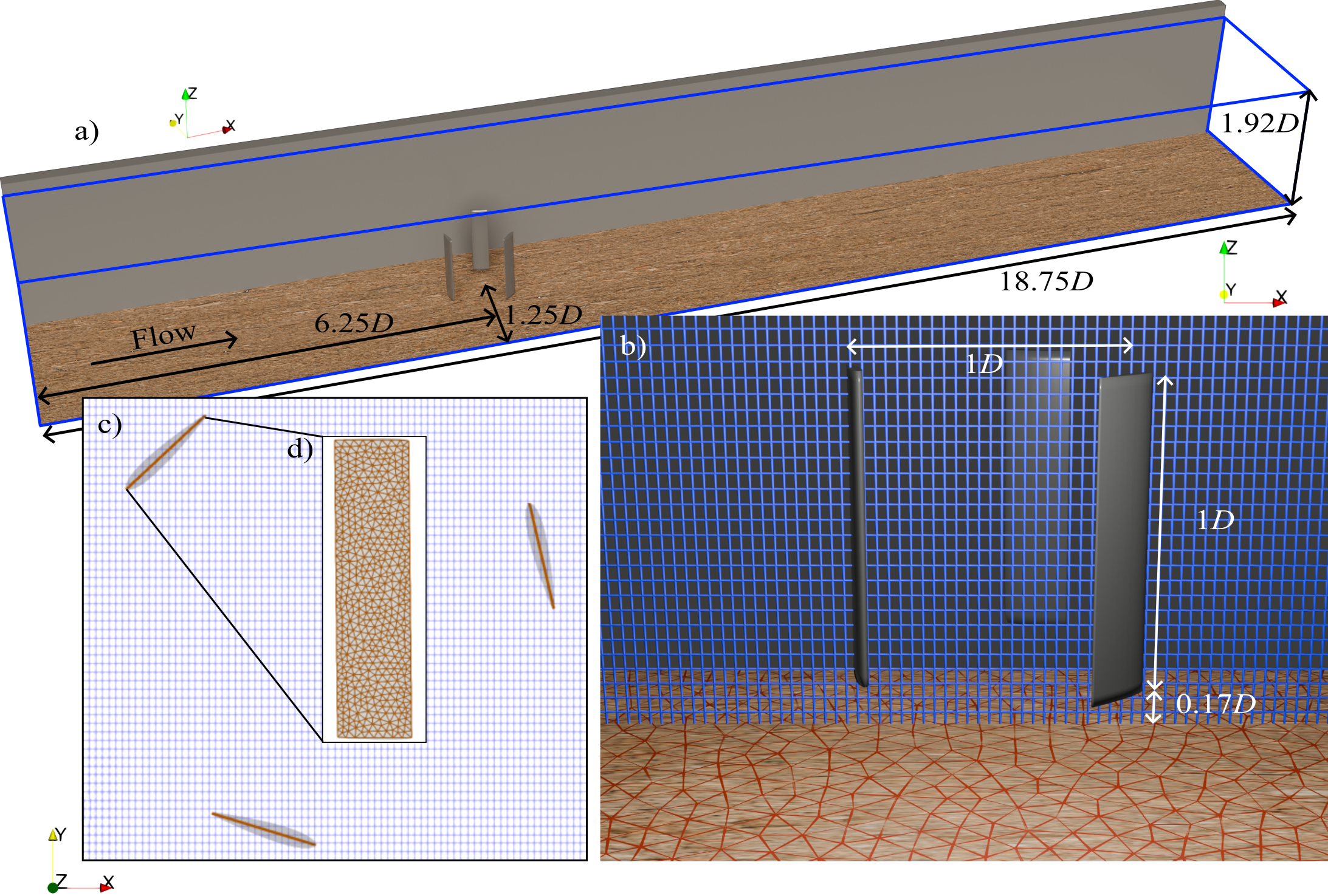} 
  \caption{Schematic of the channel and turbine in which dimensions are normalized by the rotor diameter ($D=2m$). As seen in (a), the turbine is located at $6.25D$ downstream from the channel inlet. The flow direction aligns with the positive x-axis, while the z-axis indicates the vertical direction. The channel has a total length of $18.75D$, a width of $2.5D$, and a flow depth of $1.92D$. In (b), the turbine blade dimensions are shown along with the computational grid, where the blue mesh represents the structured grid used to discretize the flow field, and the brown unstructured triangular cells discretize the bed. For visual clarity, the former and latter grid systems are coarsened by a factor of $10$ and $5$, respectively. (c) illustrates the top view of the actuator surface embedded within the blade geometry for clarity, and (d) displays the unstructured triangular mesh cells defining the actuator surface.}
  \label{Fig:1}
  \label{fig:1}
\end{figure}

\noindent This section outlines the numerical simulation setup by detailing flow conditions, turbine specifications, channel properties, and sediment particle properties. The simulated domain of the channel is $5m$ wide, $37.5m$ long, and $3.84m$ high. A utility-scale three-bladed H-Darrieus vertical-axis turbine, using a NACA$0015$ hydrofoil \cite{[29],[30]}, is positioned at the centerline and $6.25D$ downstream from the channel's inlet (\prettyref{Fig:1}(a)). The turbine has a diameter of $2m$, chord length of $0.5m$, and blade height of $2m$ (\prettyref{Fig:1}(b)), yielding a geometric solidity of $\sigma = 0.24$~\cite{[155]}. As noted by \citet{[155]}, the effects of turbine struts and the central shaft on both the flow field and bed morphodynamics are negligible. Consequently, only the turbine blades are modeled in this study, allowing for more computationally efficient simulations. Simulations are conducted at a bulk velocity of $U_{\infty} = 1.5m/s$, corresponding to a Reynolds number of $Re = 3 \times 10^6$.

The flow domain is discretized using a uniform grid with $1301 \times 141 \times 93$ nodes respectively in the streamwise, spanwise, and vertical directions. This leads to a total of approximately $17$ million background grid nodes, with an approximate spatial resolution of $0.015D$. The minimum grid spacing in the vertical direction corresponds to approximately $900$ inner wall units. A non-dimensional time step of $\Delta t_* = 0.0005$ is employed, where $\Delta t_* = \Delta t U_{\infty}/D$. The corresponding physical time step is $\Delta t = 0.00067\mathrm{s}$, ensuring that the Courant–Friedrichs–Lewy (CFL) number remains below unity throughout the simulation.

The mean grain size in sandy riverbeds typically ranges from $0.05mm$ to $2mm$~\cite{[169]}. The considered sand bed material has a mean grain size of $d_{50} = 0.7,\mathrm{mm}$ with the porosity of $\gamma = 0.4$, and the angle of repose of $\phi = 40^{\circ}$.

Additionally, three TSR values of $1.6$, $2.0$, and $2.4$ with counterclockwise blade rotation are considered in this study. Centered around the experimentally determined optimum TSR of $1.9$ \cite{[29]}, this range is selected to examine the influence of live-bed conditions on the near-optimal performance of the VAT \cite{[155]}.

\begin{table}[H]
\centering
\begin{tabular}{p{8cm} p{3cm}}
\hline
\multicolumn{2}{c}{\textit{Hydrodynamics solver}} \\
\hline
$N_x, N_y, N_z$ & $1301 \times 141 \times 93$ \\
$\Delta x, \Delta y, \Delta z$ & $0.01D$ \\
$\Delta t_{*}$ & $0.0005$ \\
$z^+$ & $600$ \\
\hline
\multicolumn{2}{c}{\textit{Morphodynamics solver}} \\
\hline
$\Delta t_s$ & $0.05$ \\
$\Delta s$ & $0.023D$ \\
$\gamma $ & $0.41$ \\
$\rho_s (kg/m^3)$ & $2650$ \\
$\phi$ & $40^{\circ}$ \\
$d_{50}(mm)$ & $0.7$ \\
\hline
\end{tabular}
\caption{Numerical setup for the hydro- and morphodynamic solvers. The computational grid resolution is denoted by $(N_x, N_y, N_z)$ in the streamwise, spanwise, and vertical directions. Spatial steps of the flow solver $(\Delta x, \Delta y, \Delta z)$ are normalized by the rotor diameter $D$, and $\Delta s$ is used in the morphodynamic solver. The minimum wall-normal spacing is reported as $\Delta z^{+}$. Time steps are expressed in non-dimensional form as $\Delta t_{*}$ for the flow solver and $\Delta t_s$ for the morphodynamic solver. Sediment porosity is denoted by $\gamma$, the angle of repose by $\phi$, sediment particle density by $\rho_s$, and the mean grain size by $d_{50}$.}
\label{tab:1}
\end{table}

For live bed cases ($4$–$6$), the bed was initially frozen while the flow solver ran until the flow reached a statistically steady state, indicated by stabilized total kinetic energy \cite{[155]}. Then, the sediment transport module was activated to allow the live bed's evolution. To reduce computational cost, a rigid-lid assumption was used to neglect water surface fluctuations around immersed bodies \cite{[68], [69]}.

\begin{table}[H]
\setlength{\tabcolsep}{5pt}      
\centering
\begin{tabular}{c c c c c c c}
\hline
 \textit{Test case} & \textit{Mobility} & \textit{TSR} 
\\
\hline
$1$  & Rigid & $1.6$ \\
$2$  & Rigid & $2.0$ \\
$3$  & Rigid & $2.4$ \\
$4$  & Live & $1.6$ \\
$5$  & Live & $2.0$ \\
$6$  & Live & $2.4$ \\
\hline
\end{tabular}
\caption{Test cases descriptions. Cases $1$ to $3$ are conducted under rigid-bed, and cases $4$ to $6$ are conducted under live-bed. TSR denotes tip-speed ratio.}
\label{tab:2}
\end{table}

A separate precursor simulation with periodic boundary conditions in the streamwise direction was run in a rigid-bed channel without any turbine to generate fully developed turbulence identified by a plateau in total kinetic energy~\cite{[155]}. The instantaneous flow field at mid-length was extracted as the inlet condition for test cases, and a Neumann boundary condition was applied at the outlet.

Simulations were performed on a $19$-core AMD Epyc Linux cluster. Rigid-bed cases required approximately $4400$ CPU hours to achieve flow convergence, whereas live-bed cases required about $6200$ CPU hours so that the live bed could reach a state of equilibrium. Compared with turbine-resolving simulations \cite{[155]}, the ASM reduced computational cost by a factor of roughly $24.5$ for rigid-bed cases and $26.6$ for live-bed cases.

\section{Results and discussions}
\label{sec:4}
\noindent This section presents the analysis of hydrodynamic and morphodynamic results, starting with instantaneous and time-averaged flow over a rigid bed, followed by live-bed simulation results. It concludes with an evaluation of turbine performance to assess the impact of sediment dynamics on VAT efficiency. 

\subsection{Wake flow under rigid-bed conditions}
\label{subsec:Wake analysis under live bed conditions}


\begin{figure} [H]
  \includegraphics[width=1\textwidth]{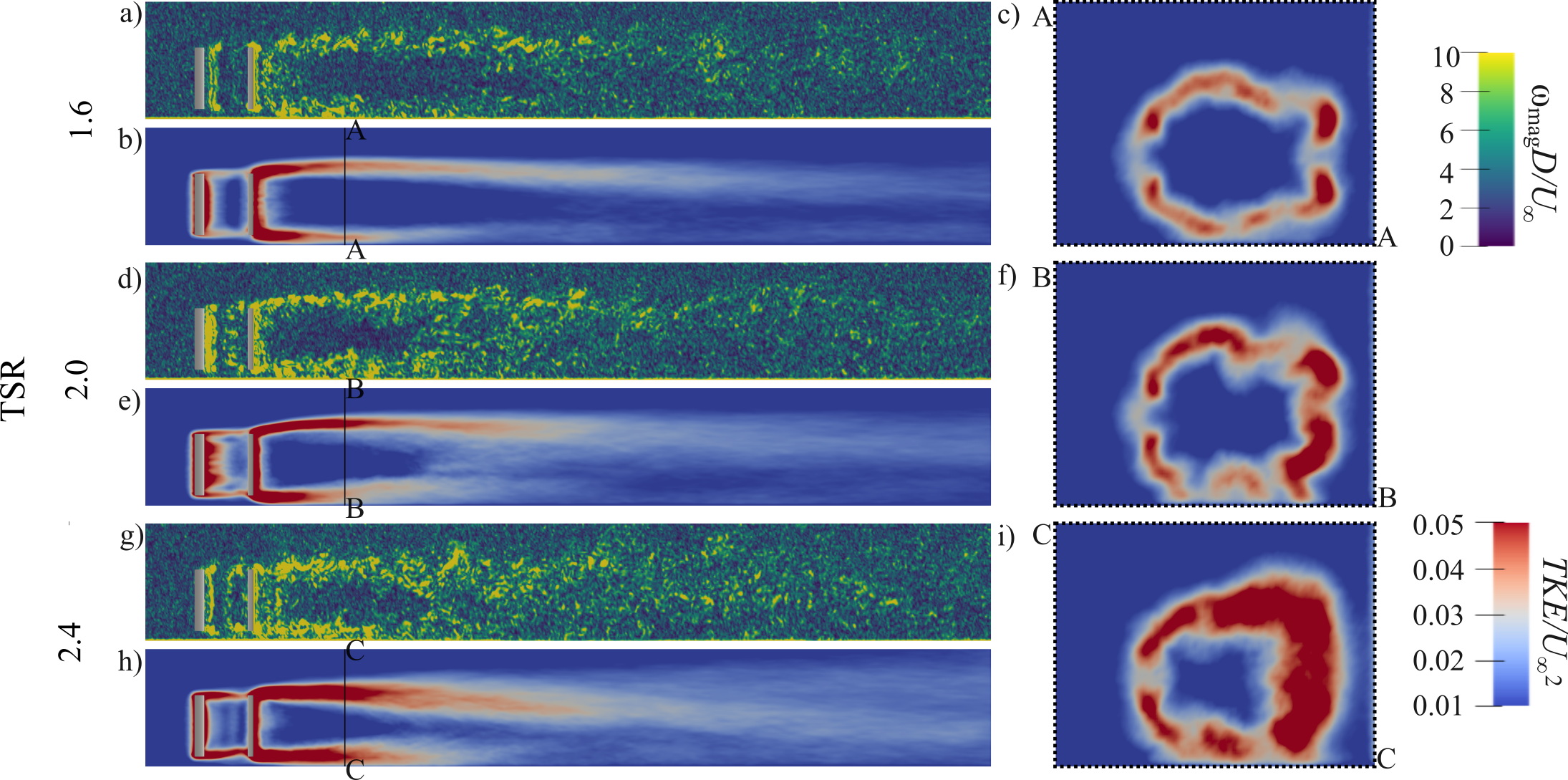} 
  \caption{Color maps of hydrodynamic results under rigid-bed conditions (cases 1–3). Panels (a), (d), and (g) show the nondimensional instantaneous vorticity magnitude from the side view at the channel centerline, corresponding to TSR = 1.6, 2.0, and 2.4, respectively. Panels (b), (e), and (h) present the TKE from the side view at the channel centerline for TSR = 1.6, 2.0, and 2.4, respectively. Panels (c), (f), and (i) represent the TKE from the cross-plane located 2D downstream of the turbine for TSR = 1.6, 2.0, and 2.4, respectively. Flow direction is from left to right.}
  \label{Fig:2}
  \label{fig:2}
\end{figure}

\noindent \prettyref{Fig:2}(a),(d),(g) presents side-view slices of the nondimensional instantaneous vorticity magnitude under rigid-bed conditions, captured at the channel centerline. As seen in all cases, the pronounced vorticity and flow separation around the blades are characteristic indicators of dynamic stall \cite{[120]}. At TSR $=1.6$, the wake exhibits relatively weak flow structures and a considerably weak shear layer, which extends into the near wake but becomes attenuated in the mid-field and almost vanishes in the far-field. As TSR increases to $2.0$ and $2.4$, stronger blade–flow and blade-to-blade interactions \cite{[46]} lead to the development of more intense shear layers in the near wake, accompanied by energetic regions shifting closer to the turbine \cite{[155]}. This highlights the correlation between TSR, turbulence production, and near-wake energy content, in agreement with previous studies \cite{[46]}. Moreover, with higher TSR, the intensified wake structures are sustained further downstream, extending their influence into the far-field. Importantly, increasing TSR also enhances turbulence within two additional regions: the near-bed zone and the interior of the rotor. These localized effects directly influence bed morphology around the turbine as well as overall turbine performance.  


\prettyref{fig:2}(b),(e),(h) presents color maps of TKE along the longitudinal section at the channel centerline. With increasing TSR, regions of elevated TKE become more concentrated around the mid-depth and shift closer to the turbine and the near-wake. As noted by \citet{[117]}, this trend reflects the amplification of local turbulence induced by higher blade rotational speeds. Given the established link between TKE and sediment erosion \cite{[155]}, such distributions have important implications for live-bed channels, which will be further examined in the following sections of this study.

\prettyref{fig:2}(c),(f),(i) presents cross-plane color maps of normalized TKE at $2D$ downstream of the turbine. For all TSRs, turbulence is concentrated near the blade edges, while the centerline and mid-depth remain comparatively lower. With increasing TSR, these regions intensify and expand toward the channel center, driven by stronger flow–blade and blade-to-blade interactions \cite{[155],[46]}. Moreover, the turbulence distribution also becomes increasingly asymmetric toward the upper sidewall, reflecting the effect of counterclockwise blade rotation. These features have important implications for turbine farms, where cumulative turbulence and wake interactions influence array performance and contribute to fatigue loading on individual turbines.

\begin{figure} [H]
  \includegraphics[width=1\textwidth]{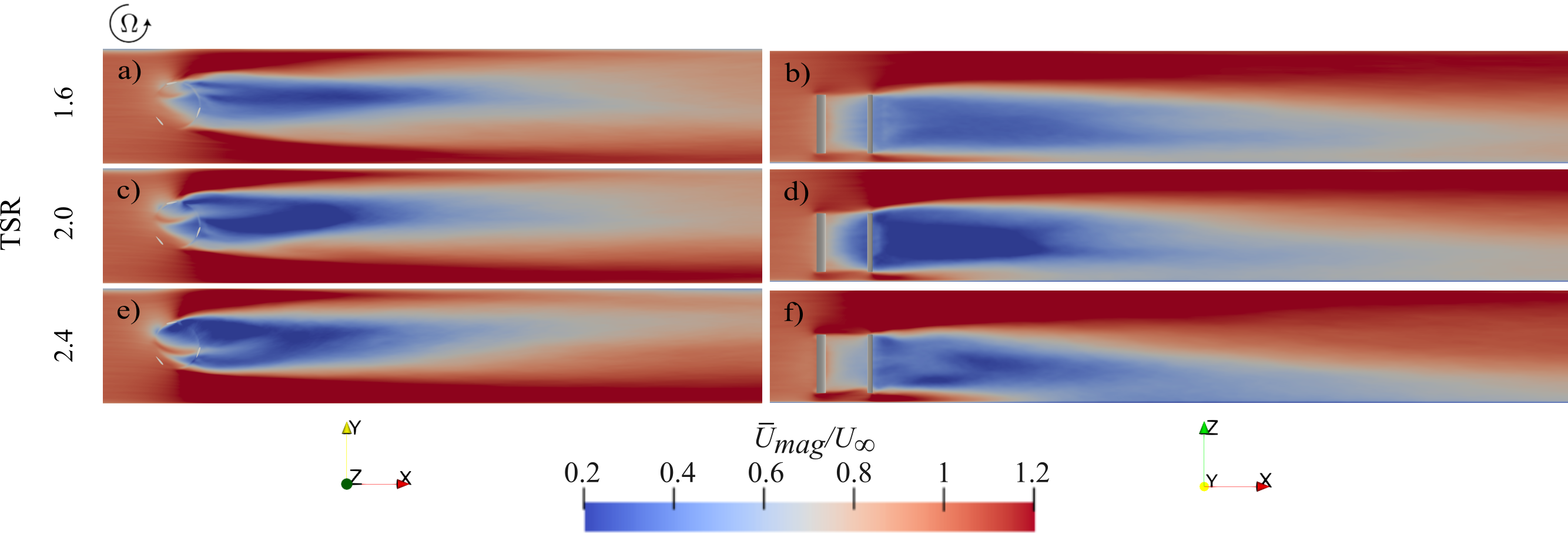} 
  \caption{Color maps of hydrodynamic results under rigid-bed conditions (i.e., cases $1$ to $3$). (a), (c), and (e) depict the nondimensional mean velocity magnitude from the top view at the mid-depth elevation of the turbine blades. (b), (d), and (f) shows the nondimensional mean velocity magnitude from the side view at the channel's centerline. The first, second, and third rows correspond to TSR $= 1.6$, $2.0$, and $2.4$, respectively. The flow is from left to right.}
  \label{Fig:3}
  \label{fig:3}
\end{figure}

\prettyref{fig:3}(a),(c),(e) presents top-view color maps of the nondimensional, time-averaged velocity magnitude ($\overline{U}_{mag}/U_{\infty}$) at blade mid-depth under rigid bed conditions. As TSR increases, the wake intensifies and becomes more confined to the near field and rotor regions (see, for example, \prettyref{fig:3}(a) and (e)). This concentration of energy reflects stronger near-field turbulence, which elevates bed shear stress and is expected to enhance sediment transport in the wake.
In the far field, the wake is wider at lower TSR, stretching downstream and covering almost the entire channel width. More specifically, at lower TSR, the broad downstream wake appears to be suppressed by the channel side walls. This might be considerable in natural riverine environments where turbines are deployed in farms, potentially impacting the power production of downstream turbines.

\prettyref{fig:3}(b),(d),(f) presents side-view color maps of the longitudinal plane of nondimensional, time-averaged velocity magnitude ($\overline{U}_{mag}/U_{\infty}$) along the channel centerline under rigid bed conditions. With increasing TSR, a distinct high-momentum region forms beneath the turbine, suggesting intensified sediment particle motion in this zone. Furthermore, the turbulent wake core progressively shifts downward in both the near and far field, aligning more closely with the bed over a longer downstream extent. This behavior indicates an increased potential for sediment transport across these regions under live-bed conditions.


\subsection{Wake flow under live-bed conditions}
\label{subsec:Wake analysis under live bed conditions}

\noindent Herein, we examine the hydrodynamic results from coupled hydro- and morphodynamic simulations conducted under live-bed conditions. Initially, the bed is held fixed until the flow reaches statistical convergence. The simulation then continues over a flat, deformable bed. As the turbulent flow interacts with the live bed, the sediment particle movement reshapes the bed until a dynamic equilibrium is achieved, characterized by stable sand wave migration as well as by nearly constant maximum scour depths and sand bar heights. Once equilibrium is reached, the bed geometry is frozen, and the flow solver is reactivated to compute time-averaged turbulence statistics.

\begin{figure} [H]
  \includegraphics[width=1\textwidth]{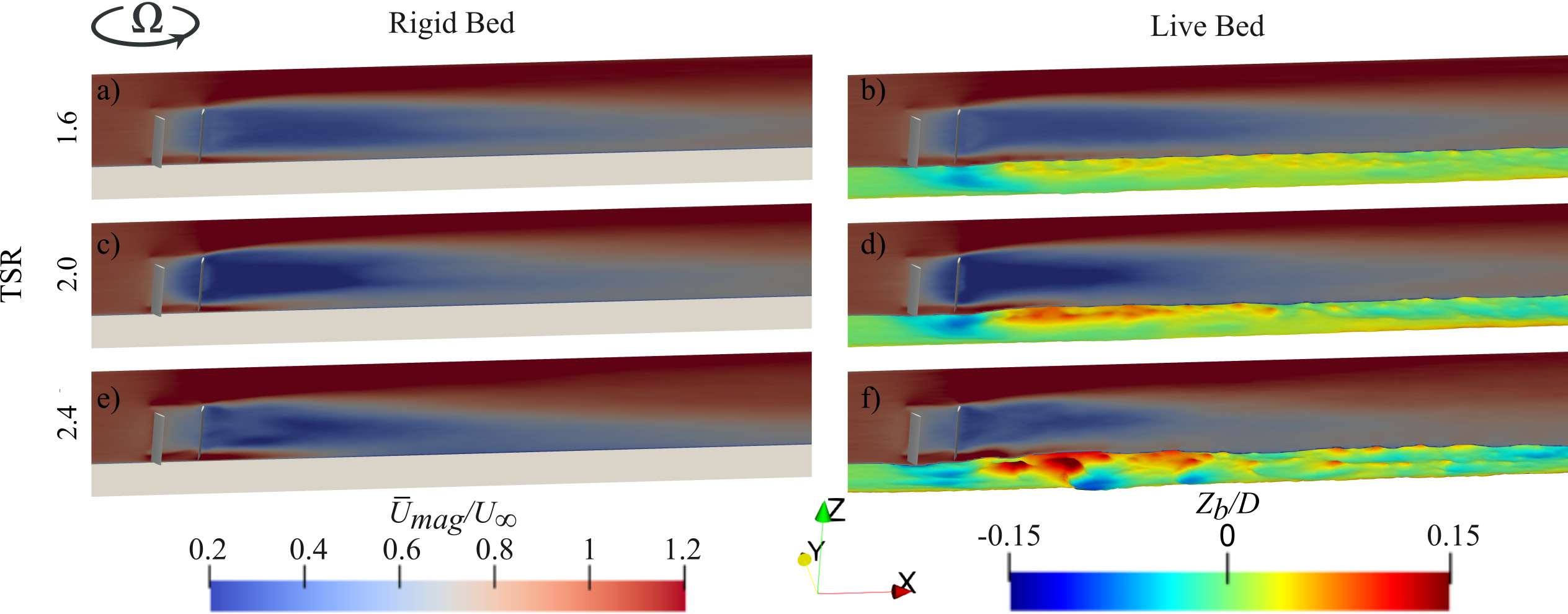} 
  \caption{Color maps of mean streamwise velocity magnitude normalized by the bulk velocity ($=1.5 m/s$) under rigid-bed (left column) and live-bed (right column) conditions, shown over vertical planes along the centerline of the channel. The live-bed results are presented together with the corresponding bed elevation color maps at the equilibrium state. The first, second, and third rows correspond to TSR = 1.6, 2.0, and 2.4, respectively. Flow direction is from left to right.}
  \label{Fig:4}
  \label{fig:4}
\end{figure}

In \prettyref{Fig:4}, we present the dimensionless time-averaged velocity magnitude ($\overline{U}_{mag}/U_{\infty}$) at the dynamic equilibrium state of the live bed for different TSRs, shown alongside the rigid-bed cases for more precise comparison. Under live-bed conditions (\prettyref{Fig:4}(b),(d),(f)), the near-bed flow momentum around the turbine is slightly reduced as a result of bed deformation. Moving upwards from the bed, the influence of live-bed effects gradually diminishes, and at approximately $0.3D$ above the deformed bed, the velocity field closely resembles that of the rigid-bed configuration. These findings are consistent with the results reported by \citet{[155]}. Moreover, the near-wake results show that under live-bed conditions, the turbine wake flow is significantly shorter and the velocity magnitude recovers more quickly compared to rigid-bed scenarios. This enhanced recovery is primarily due to bed deformations, especially scour near the turbine base and immediately downstream deposition, which generate localized turbulence and a near-bed jet flow. This jet injects momentum into the wake core, accelerating flow recovery and redistributing energy \cite{[155]}. 

\begin{figure} [H]
  \includegraphics[width=1\textwidth]{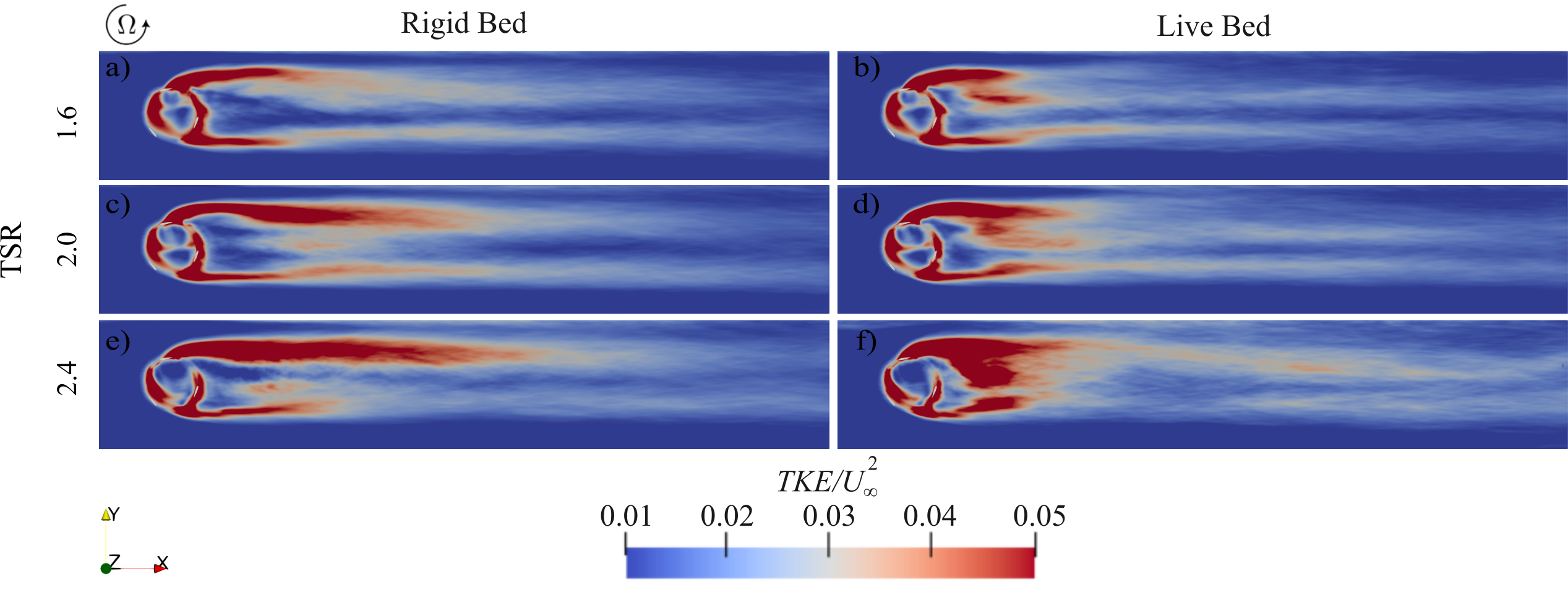} 
  \caption{Color maps of the computed TKE normalized by $U_{\infty}^2$ at an elevation of $z=0.35D$ above the bed and under rigid-bed (left column) and live-bed (right column) conditions. The first, second, and third rows correspond to TSR = 1.6, 2.0, and 2.4, respectively. Flow direction is from left to right.}
  \label{Fig:5}
  \label{fig:5}
\end{figure}

\prettyref{fig:5} shows top-view color maps of TKE, normalized by $U_{\infty}^2$, under both rigid- and live-bed conditions, captured at $0.35D$ above the bed, that is, slightly above the highest deposition observed under the live-bed case at TSR$=2.4$. Under rigid-bed conditions (\prettyref{Fig:5}(a), (c), and (e)), increasing TSR yields an asymmetric distribution of elevated turbulence, with higher levels along the upper part of the channel extending farther downstream. Under live-bed conditions (\prettyref{Fig:5}(b), (d), and (f)), however, turbulence shifts closer to the turbine, rendering the near-wake region more turbulent. However, the elevated turbulence under live-bed conditions, particularly at higher TSR, decays more rapidly than that under rigid-bed conditions (see \prettyref{Fig:5}(e) and (f)). Additionally, the effect of the live bed on the turbulent flow gradually diminishes with height.

\begin{figure} [H]
  \includegraphics[width=1\textwidth]{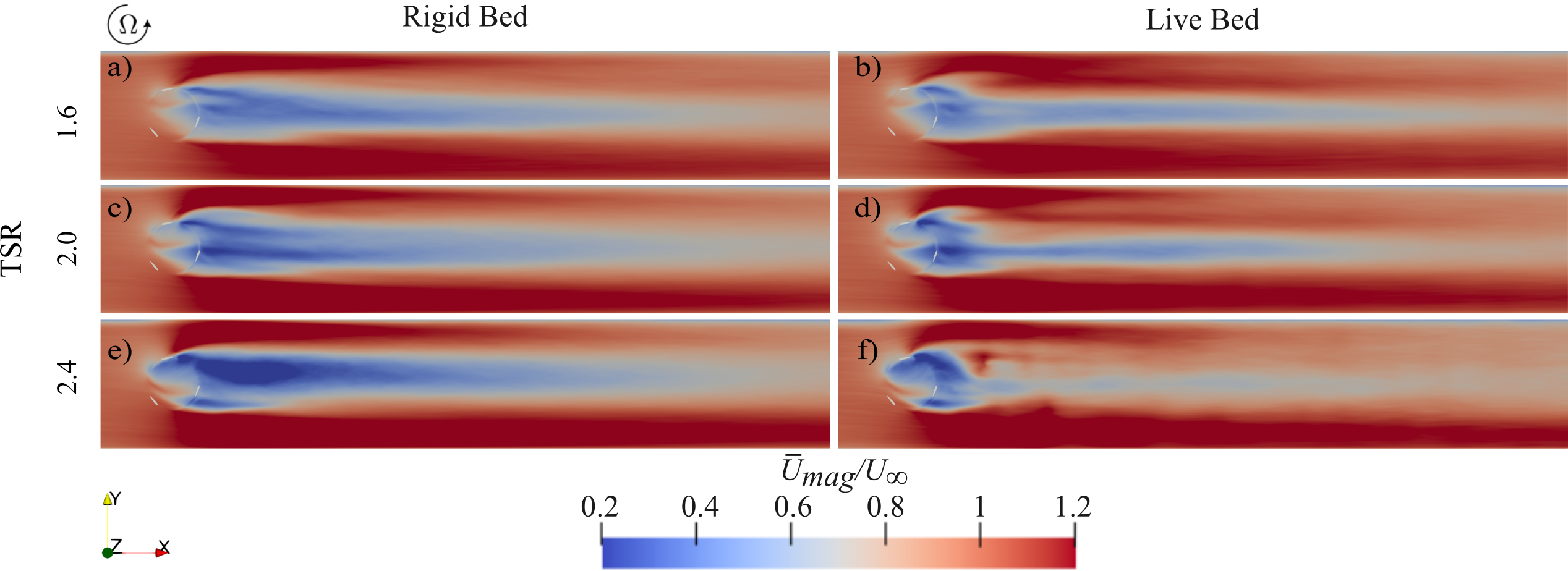} 
  \caption{Color maps of the computed mean velocity magnitude normalized by the bulk velocity ($=1.5 m/s$) at an elevation of $z=0.35D$ above the bed and under rigid-bed (left column) and live-bed (right column) conditions. The first, second, and third rows correspond to TSR = 1.6, 2.0, and 2.4, respectively. Flow direction is from left to right.}
  \label{Fig:6}
  \label{fig:6}
\end{figure}

\prettyref{fig:6} illustrates the normalized time-averaged velocity magnitude under both rigid- and live-bed cases, captured at $0.35D$ above the bed. As seen under rigid-bed conditions (\prettyref{Fig:6}(a), (c), and (e)), increasing TSR produces a highly intensified momentum-deficit region in the near wake. This high-deficit region is slightly asymmetric. More specifically, a larger velocity deficit appears along the upper part of the channel and persists farther downstream, which is aligned with previous findings \cite{[155]}.
Under live-bed conditions (\prettyref{Fig:6}(b),(d),(f)), sediment particles transported from beneath the turbine and deposited in the near field deforms the bed and generates bed-induced jet flows that penetrate the wake core, leading to faster velocity recovery through the mixing process. At TSR $=1.6$ (\prettyref{Fig:6}(a) and (b)), this effect is more apparent in the near field and diminishes farther downstream. As TSR increases, however, the effect becomes more pronounced in both the near and far fields. In other words, under live-bed conditions, increasing TSR yields faster wake recovery due to higher downstream sand bars and deeper erosion beneath the turbine, which generate stronger jet flows that inject additional momentum into the downstream wake core, consistent with previous findings \cite{[155]}.

\begin{figure} [H]
  \includegraphics[width=1\textwidth]{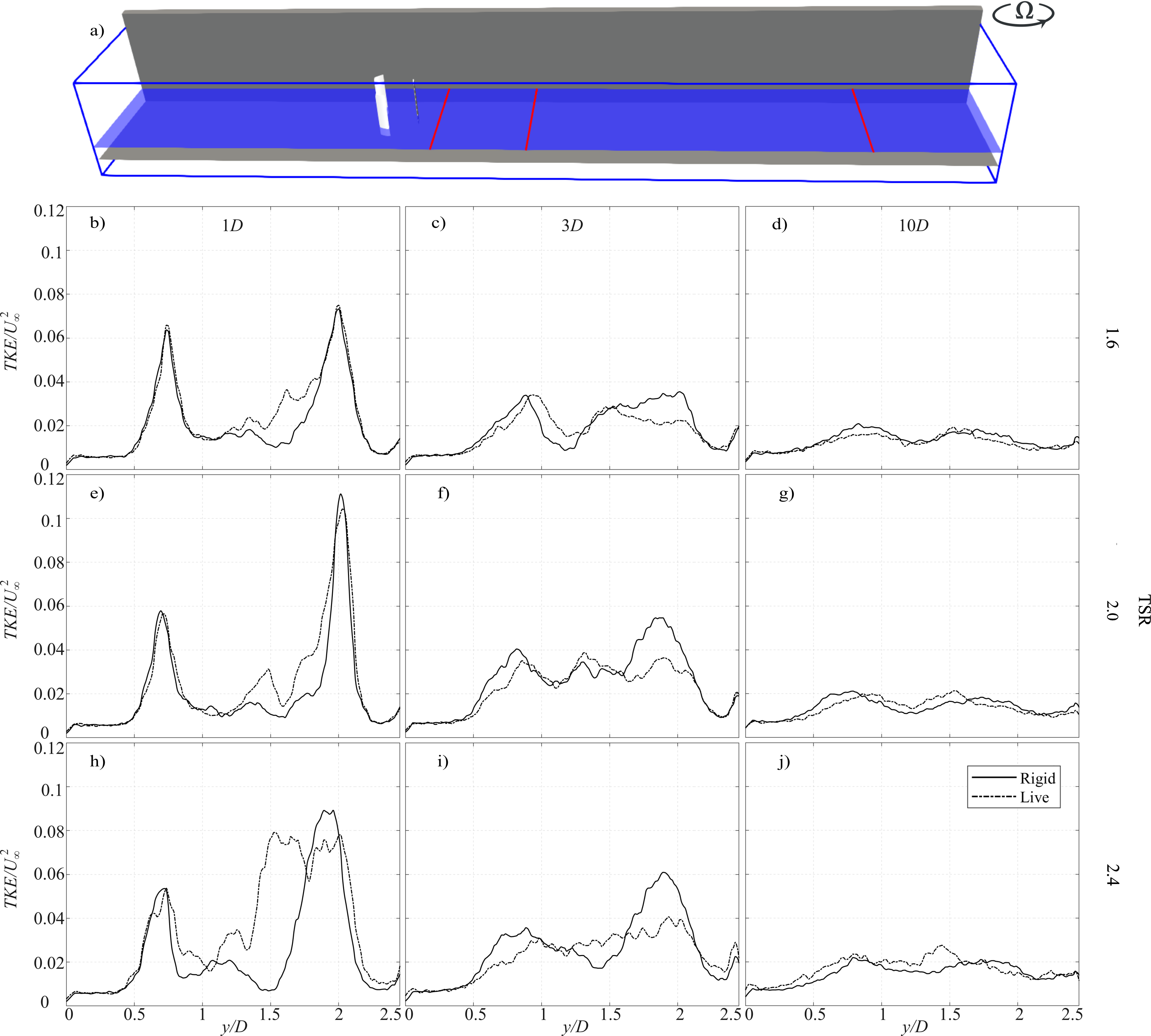} 
  \caption{Cross-stream TKE profiles at $0.35D$ above the bed at $1D$, $3D$, and $10D$ downstream of the turbine, normalized by $U_{\infty}^2$, under both rigid and live bed conditions. The first, second, and third rows correspond to TSR $= 1.6$, $2.0$, and $2.4$, respectively.}
  \label{Fig:7}
  \label{fig:7}
\end{figure}

\begin{figure} [H]
  \includegraphics[width=1\textwidth]{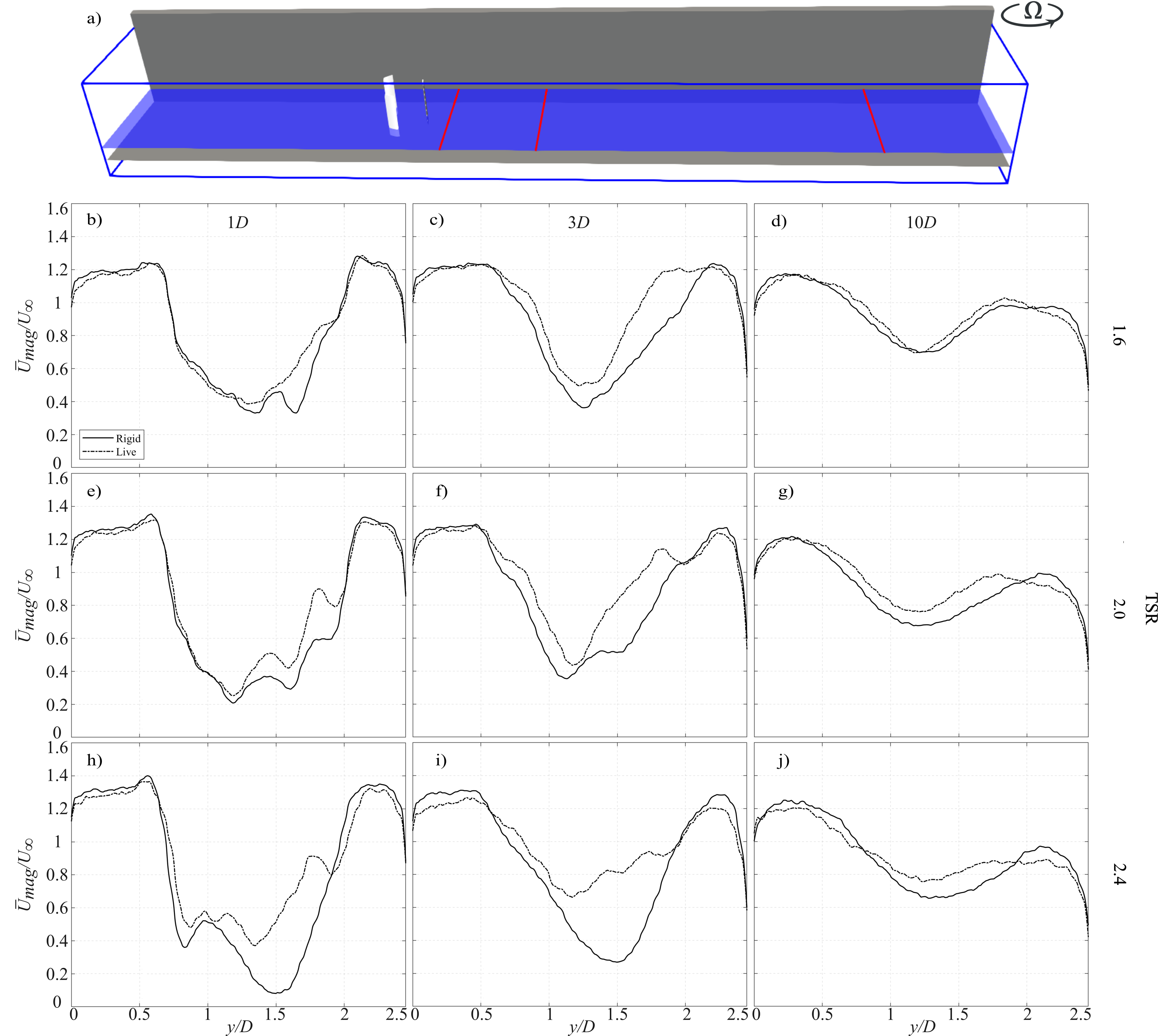} 
  \caption{Cross-stream velocity magnitude profiles at $0.35D$ above the bed at $1D$, $3D$, and $10D$ downstream of the turbine, normalized by the bulk velocity ($=1.5 m/s$), under both rigid and live bed conditions. The first, second, and third rows correspond to TSR $= 1.6$, $2.0$, and $2.4$, respectively.}
  \label{Fig:8}
  \label{fig:8}
\end{figure}

For a quantitative assessment of the turbulent flow under both rigid- and live-bed conditions, \prettyref{Fig:7} and \prettyref{Fig:8} correspond to the TKE and velocity magnitude profiles, respectively, extracted at $0.35D$ above the bed along the spanwise direction. The profiles are sampled at three downstream locations, $x/D=1$, $x/D=3$, and $x/D=10$, corresponding to the near, mid, and far field. We begin with the analysis of TKE. Under rigid-bed conditions at $x/D=1$ (\prettyref{Fig:7}(b), (e), and (h)), increasing TSR generally elevates TKE. However, in the lower part of the channel (i.e., $y/D\lesssim0.75$), the differences across cases remain below $3\%$ and are therefore negligible. Approaching the centerline, the inter-case difference grows to about $5\%$, with the highest TSR exhibiting the largest turbulence. From the channel centerline toward the upper region, where the blades encounter the incoming flow and turbulence intensifies, the inter-case differences grow progressively, reaching their maximum near the edge of the shear layer. More specifically, at $y/D\simeq1.75$, TKE increases from $0.022$ for TSR $=1.6$ to $0.058$ for TSR $=2.4$, which is about a $163\%$ increase. For $1.95\lesssim y/D\lesssim2.1$, TKE decreases for all cases except TSR $=2.0$, which peaks near $y/D\simeq2.0$ at $0.11$. Finally, in the upper part of the channel ($y/D\gtrsim2.1$), blade effects on the flow almost diminish, and TKE declines at nearly the same rate across all cases.
Under live-bed conditions, the bed deformations increase TKE in all cases. This probably stems from the deformed bed beneath and in front of the turbine. 
The elevated TKE is primarily concentrated in $1.2\lesssim y/D\lesssim1.75$, the region most influenced by blade rotation, where the associated bed deformations further amplify the turbulence. Additionally, the difference in TKE between the live- and rigid-bed cases increases with increasing TSR. More specifically, under live-bed conditions, TKE increases on average by $14.4\%$, $19.2\%$, and $43.1\%$ for $TSR=1.6$, $2.0$, and $2.4$, respectively.

At $x/D = 3$ (\prettyref{Fig:7}(c), (f), and (i)), TKE increases under live-bed conditions in the channel centerline region ($0.95\lesssim y/D\lesssim1.5$). This rise is attributed to tall sand bars downstream of the turbine, which manipulate momentum and enhance turbulence in this area. In contrast, outside this region and closer to the sidewalls, TKE decreases under live-bed conditions. More specifically, under live-bed conditions and along the centerline, TKE increases by $66.2\%$, $48.1\%$, and $58.8\%$ for TSR $= 1.6$, $2.0$, and $2.4$, respectively, compared to that under rigid-bed conditions. Conversely, near the sidewalls, it decreases by $42.9\%$, $38.8\%$, and $96.7\%$ for TSR $= 1.6$, $2.0$, and $2.4$, respectively. From a hydrodynamic perspective, in wider farm configurations, reduced TKE within approximately $1D$ above and below the turbine can significantly alter flow structures around downstream turbines. Although structural implications are beyond the scope of this work, these results suggest that reduced TKE near the sidewalls under live-bed conditions can help to mitigate erosion in narrow riverine environments, supporting turbine installation by reducing bank instability.
At $x/D=10$ (\prettyref{Fig:7}(d), (g), and (j)), the chaotic and turbulent flow observed in the near- and mid-field is somewhat moderated, and the live-bed influence on TKE in the far field remains relatively small, with an average effect of less than $3\%$. Similar to the mid-field, under live-bed conditions, TKE increases around the channel centerline, while it decreases outside this region and closer to the sidewalls. An exception occurs at TSR $=2.4$, where TKE increases near the lower sidewall under live-bed conditions. This is likely related to bed deformations induced at higher TSR, which extend further downstream and alter the local flow field.


We then proceed with the analysis of velocity magnitude in \prettyref{Fig:8}.
Under rigid-bed conditions at $x/D=1$ (\prettyref{Fig:8}(b), (e), and (h)), wake symmetry is influenced by TSR. At lower TSR, the downstream wake remains more symmetric, whereas higher TSR values skew the wake upwards, reflecting the counterclockwise blade rotation. Moreover, increasing TSR reduces the minimum near-wake velocity around the channel centerline to $0.33$, $0.20$, and $0.07$ for TSR $=1.6$, $2.0$, and $2.4$, respectively. In contrast, velocities near the sidewalls ($y/D \lesssim 0.7$ and $y/D \gtrsim 2.0$) remain nearly unchanged, indicating the faster wake recovery in the spanwise direction with increasing TSR.
Under live-bed conditions (\prettyref{Fig:8}(c), (f), and (i)), the near wake recovers considerably in all TSRs. This recovery is driven by jet flows induced by near-wake bed deformation, which inject momentum into the wake core and modify its structure. The effect intensifies with increasing TSR, as taller sand bars amplify the jet and its influence on the wake. More specifically, the minimum wake velocity under live-bed conditions is $0.39$, $0.25$, and $0.37$ for TSR$=1.6$, $2.0$, and $2.4$, respectively. Near the sidewalls, the influence of the live bed diminishes, and the flow pattern closely resembles that observed over a rigid bed.

Further downstream at $x/D=3$ (\prettyref{Fig:8}(c), (f), and (i)), the wake deficit trend persists under both rigid- and live-bed conditions. Under the rigid bed, increasing TSR lowers the minimum near-wake velocity to $0.36$, $0.33$, and $0.27$ for TSR $=1.6$, $2.0$, and $2.4$, respectively. A comparison with values at $x/D=1$ shows that velocity recovery becomes faster as TSR increases. More specifically, from $x/D=1$ to $x/D=3$, the minimum velocity rises by about $9\%$, $65\%$, and $237\%$ corresponding to TSR $=1.6$, $2.0$, and $2.4$, respectively. Under live-bed conditions, the minimum flow velocity increases, and this effect becomes more pronounced with higher TSR. More specifically, under live bed conditions, the minimum wake velocity reaches $0.45$, $0.46$, and $0.66$ corresponding to TSR$=1.6$, $2.0$, and $2.4$, respectively. These values represent increases of $25\%$, $39\%$, and $144\%$ compared to those observed under rigid-bed conditions.
At $x/D = 10$ (\prettyref{Fig:8}(f), (g), and (j)), a substantial portion of the velocity has recovered in both rigid- and live-bed conditions. Under rigid-bed conditions, the minimum wake velocity deficit reaches $0.70$, $0.64$, and $0.63$, corresponding to TSR $=1.6$, $2.0$, and $2.4$, respectively. In comparison, under live-bed conditions, the minimum velocity increases slightly to $0.71$, $0.77$, and $0.79$ corresponding to TSR $=1.6$, $2.0$, and $2.4$, respectively. This suggests that as TSR decreases, the bed is less deformed by blade-driven turbulence, and in the far field, the influence of bed deformation on the wake flow diminishes.

\subsection{Sediment dynamics}
\label{subsec:Wake analysis under live bed conditions}

\begin{figure} [H]
  \includegraphics[width=1\textwidth]{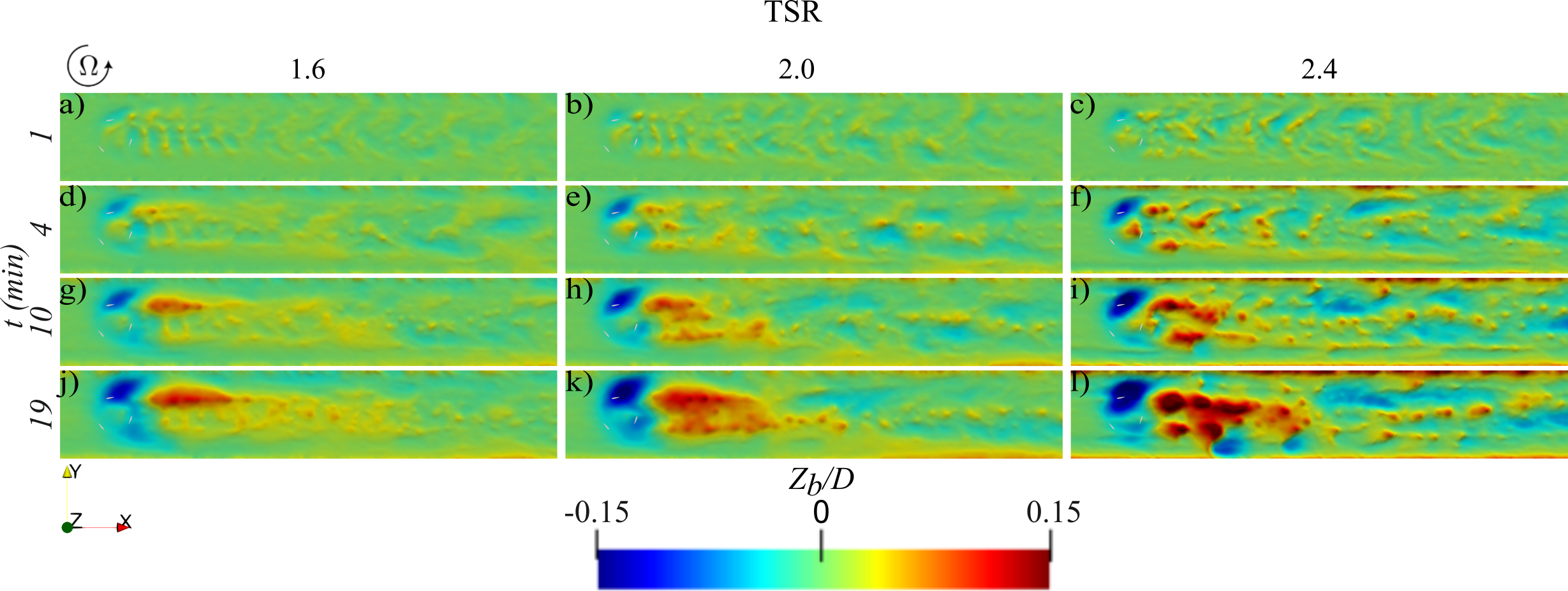} 
  \caption{Color maps of bed elevation evolution, normalized with rotor diameter ($D=2m$) for cases $4$ to $6$, captured from the top view. The first, second, third, and fourth columns correspond to bed elevations captured at $1$, $4$, $10$, and $19$ minutes of physical time of the live-bed evolution, respectively. The fourth row represents the equilibrium state of the live bed. The first, second, and third columns correspond to TSR $= 1.6$, $2.0$, and $2.4$, respectively. The flow is from left to right.}
  \label{Fig:9}
  \label{fig:9}
\end{figure}
\begin{figure} [H]
  \includegraphics[width=1\textwidth]{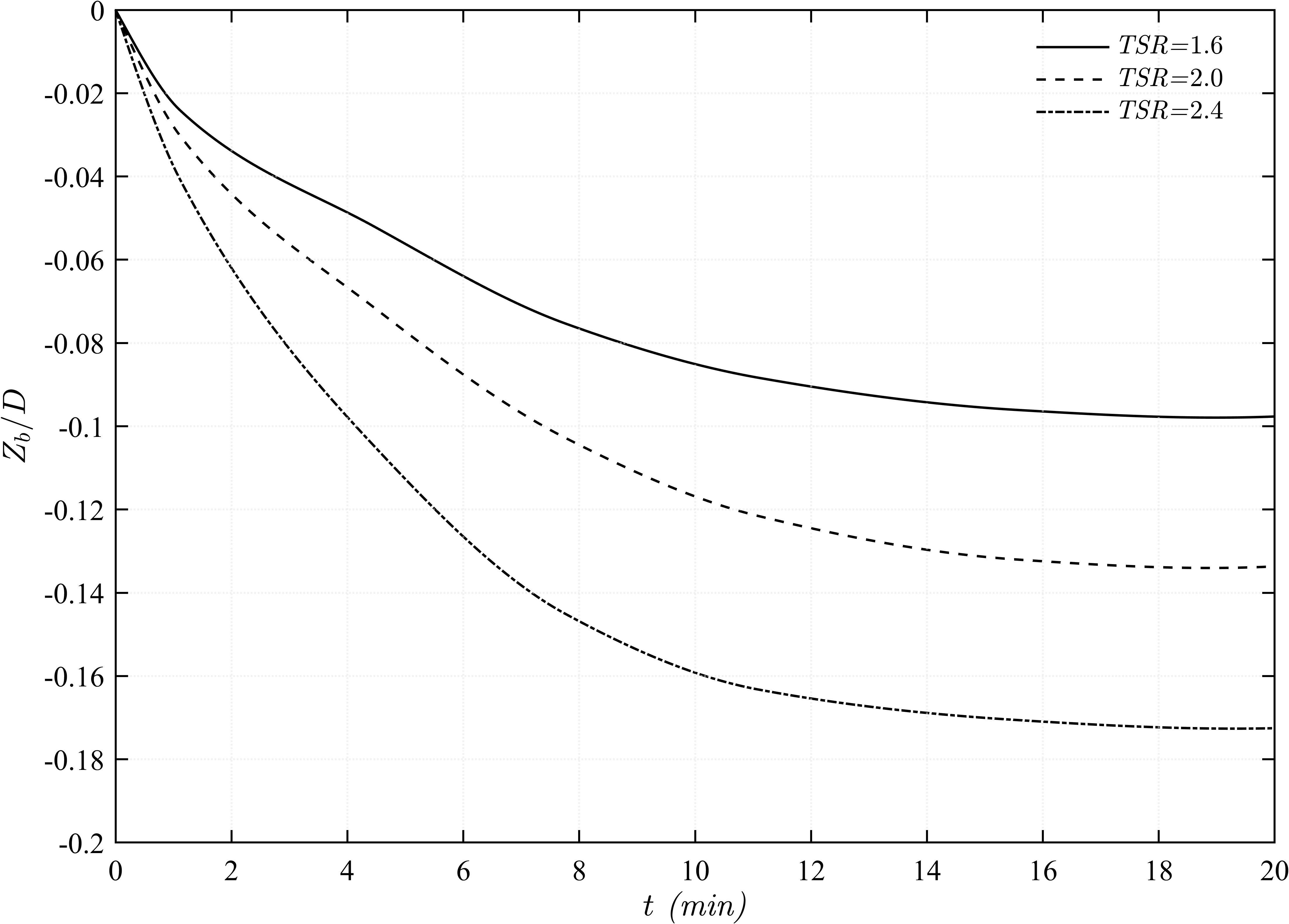} 
  \caption{Temporal evolution of maximum bed erosion across the live-bed channel for different cases.}
  \label{Fig:10}
  \label{fig:10}
\end{figure}

In \prettyref{Fig:9}, we plot the bed deformation normalized by the rotor diameter ($Z_b/D$) over time, from 1 to 19 minutes of physical time from the morphodynamic simulations. Over time, small-scale sand waves (ripples) form across the bed, grow in amplitude, and migrate downstream \cite{[155]}. As waves of different sizes develop and migrate, the near-wake portion of the bed undergoes pronounced deformation, with the formation of a dominant scour hole beneath the turbine, followed immediately downstream by sand bars. The scour depth and sand-bar deposition height increase continuously until a dynamic equilibrium is reached. More specifically, we record the maximum erosion across the channel, which for all cases occurs beneath the turbine and slightly above the channel centerline, and plot its evolution in \prettyref{Fig:10} as a function of physical time after activating the sediment-transport module. As shown in this figure, the maximum bed erosion decreases over time and reaches a plateau, indicating dynamic equilibrium \cite{[69]}. This equilibrium state was used for the subsequent bed-deformation analysis. Additionally, the morphodynamic simulations for all cases reach equilibrium at approximately 19 minutes of physical time.

As seen in the live beds' equilibrium state (\prettyref{Fig:9}(j), (k), and (l)), the rotating blades and the associated turbulence lead to considerable sediment transport in both the near and far field of the turbine. More specifically, although bed deformation is negligible upstream of the turbine, downstream sediment transport is substantial and produces sand waves migrating with varying amplitudes.
The rotating blades erode sediment beneath the turbine and carry it downstream, where the gradual decay of flow velocity within the wake promotes deposition in the near field, leading to the formation of sand bars. As the bed continues to evolve, these bars partially obstruct and trap sediment carried from upstream and beneath the turbine, leading to progressively taller bars with accelerated growth rates. Additionally, TSR plays a crucial role in these deformations. More specifically, as TSR increases, turbulence intensifies, the scour hole beneath the turbine becomes deeper and wider, and the downstream sand bars become higher and longer. At equilibrium, the scour hole depths measure approximately $-0.0979D$, $-0.1340D$, and $-0.1726D$ for TSR values of $1.6$, $2.0$, and $2.4$, respectively, while the sand bar heights reach $0.122D$, $0.143D$, and $0.240D$ for the same TSR values. Moreover, in the far field, the turbine-induced wake continues to transport sediment, with higher TSRs driving greater sediment movement farther downstream. These bed deformations are particularly important in VAT farms, as they alter the flow structure downstream and may ultimately influence the overall performance of turbine arrays.

\begin{figure} [H]
  \includegraphics[width=1\textwidth]{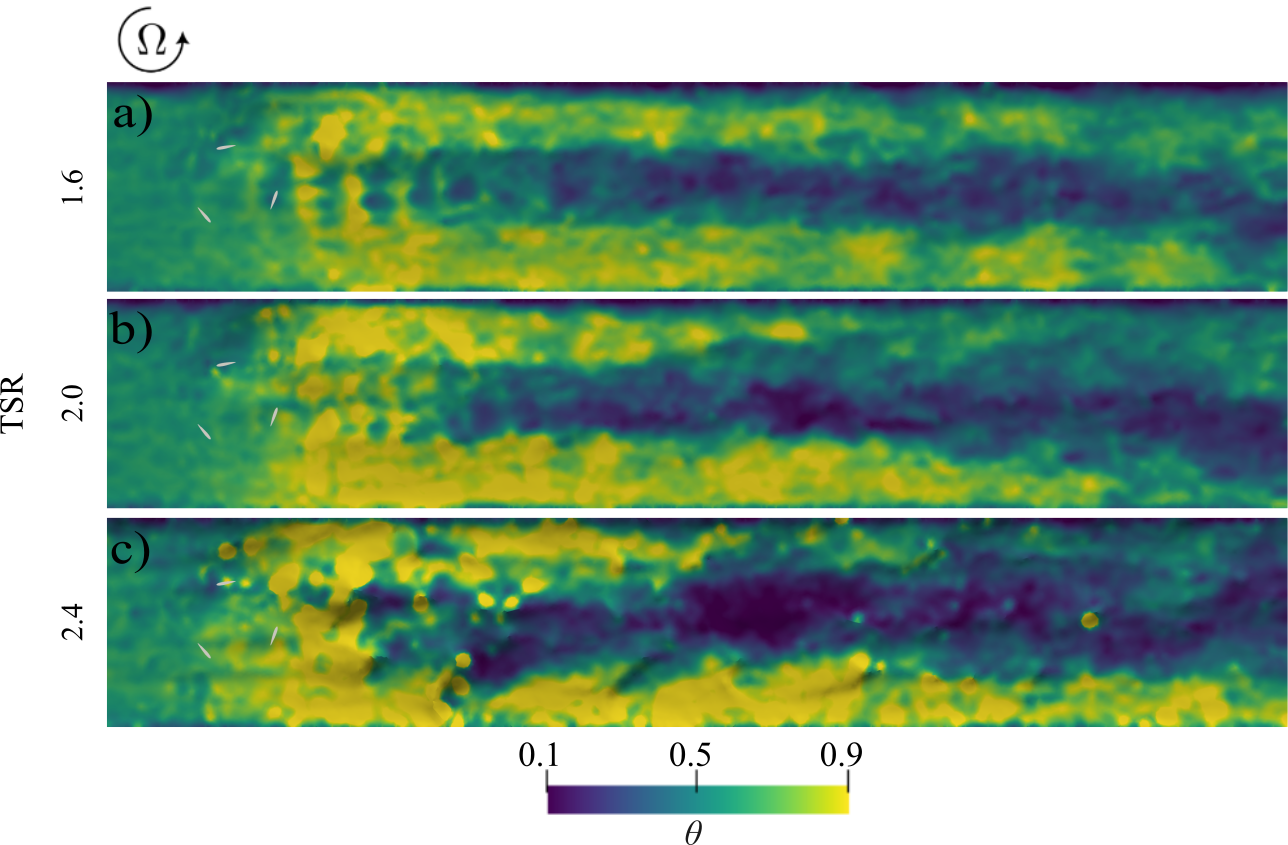} 
  \caption{Color maps of the distribution of dimensionless instantaneous bed shear stress (i.e., Shiled's parameter), captured from the $0.35D$ above the bed at the dynamic equilibrium state for cases $4$ to $6$. The first, second, and third rows correspond to TSR $= 1.6$, $2.0$, and $2.4$, respectively. The flow is from left to right.}
  \label{Fig:11}
  \label{fig:11}
\end{figure}

In \prettyref{Fig:11}, we present the distribution of the dimensionless instantaneous bed shear stress at equilibrium state, i.e., Shields parameter ($\theta$), which is defined as \cite{[105]}:
\begin{equation}
    \theta = \frac{\tau_*}{(\rho_s - \rho) g d_{50}}
    \label{eq:24}
\end{equation}

As observed, the signature of the rotating blades produces a strong shear layer downstream of the turbine that extends farther downstream. A zone of elevated bed shear stress develops beneath and immediately downstream of the turbine as a result of blade rotation, while an additional high-shear region forms near the sidewall, reflecting the interaction of turbulent flow with the solid boundary.
In contrast, farther downstream, the lower-shear region around the channel centerline, shows the opposite behavior with respect to bed elevation. Moreover, a comparison of \prettyref{Fig:9} and \prettyref{Fig:11} highlights the strong relationship between major bed deformations and the Shields parameter, $\theta$. Regions exhibiting large variations in $\theta$ correspond to significant divergences in sediment flux, which ultimately drive pronounced bed deformation. In particular, sediment entrained within the relatively high-shear zone beneath the turbine is carried downstream, and as bed shear stress diminishes farther downstream, this sediment is deposited to form sand bars (see, for example, \prettyref{Fig:9}(c) and \prettyref{Fig:11}(l)). Moreover, although the shear layer extends downstream at all TSRs, its intensity grows with increasing TSR. For example, at TSR $=2.4$ (\prettyref{Fig:11}(c)), the region of elevated bed shear stress not only stretches farther downstream but also exhibits higher intensity, reaching closer to the lower sidewall, where it mobilizes sediment particles and generates substantial bed deformations that persist to the end of the channel.

\subsection{Turbine performance}

\noindent Herein, we examine the turbine’s performance under both rigid- and live-bed conditions. To this end, the mean power coefficient is defined as the ratio of the average extracted power to the theoretical maximum power available in the flow, expressed as follows \cite{[27], [40], [69], [155]}:

\begin{equation}
C_p = \frac{\overline{P}}{P_{max}}
\end{equation}

\noindent Here, the average extracted power is obtained from the product of the mean torque applied to the shaft and the turbine’s angular velocity as follows \cite{[27], [40], [69], [126], [155]}:

\begin{equation}
\overline{P} = \overline{T} \times \Omega
\end{equation}

\noindent The maximum available power is given by:

\begin{equation}
P_{max} = 0.5 \rho U_{\infty}^3 A
\end{equation}

\noindent where $\rho$ is the fluid density, $U_{\infty}$ is the bulk velocity, and $A$ is the turbine’s swept area. According to Betz’s theory, the upper limit of $C_p$ is $16/27$ \cite{[127]}.

\begin{figure} [H]
  \includegraphics[width=1\textwidth]{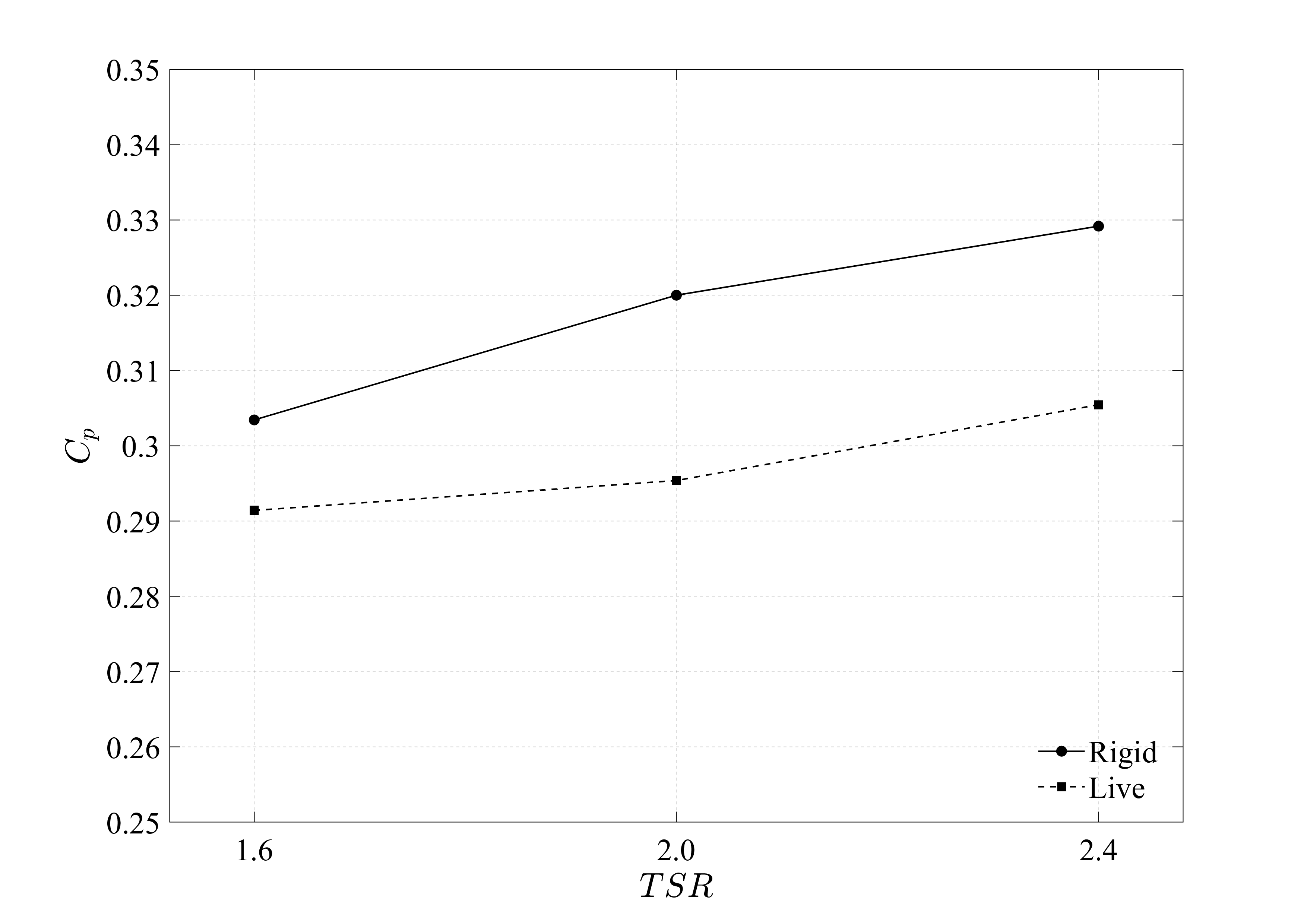} 
  \caption{The mean power coefficient of the single utility-scale turbine under the rigid- and live-bed conditions.}
  \label{Fig:12}
  \label{fig:12}
\end{figure}

As shown in \prettyref{Fig:12}, the mean power coefficient is presented for all cases under both rigid- and live-bed conditions. Under rigid-bed conditions, the power coefficient is observed to decline with decreasing TSR. This behavior aligns with earlier findings by \citet{[27]}, which indicate that operation below the optimal TSR is accompanied by intensified dynamic stall effects, ultimately resulting in reduced turbine efficiency. More specifically, the power coefficient decreases progressively with TSR, taking values of $0.329$, $0.320$, and $0.303$ for TSR $=2.4$, $2.0$, and $1.6$, respectively. This behavior is consistent with the results reported by \citet{[155]}.

Under live-bed conditions, the mean power coefficient is reduced at all TSRs compared to those observed under the rigid-bed conditions. Moreover, the reduction in efficiency becomes more pronounced as TSR increases. More specifically, turbine efficiency drops by $1.1\%$, $2.2\%$, and $2.5\%$ for TSR $=1.6$, $2.0$, and $2.4$, respectively, relative to the rigid-bed conditions.
Additionally, as shown in \prettyref{Fig:12}, the increasing trend of turbine performance with rising TSR under rigid-bed conditions suggests that the optimum TSR lies near $2.4$, since the incremental improvements in turbine performance become progressively smaller as TSR increases, indicating that the turbine is approaching its maximum power coefficient \cite{[27]}. Under live-bed conditions, however, this trend is disrupted, and the optimum TSR appears to shift to values greater than $2.4$. In other words, the live bed not only alters the momentum exchange around the turbine, leading to reduced efficiency, but also modifies the optimum TSR value. Nonetheless, additional cases should be examined to confirm the precise value of the optimum TSR \cite{[155]}. Importantly, although turbines in real-world environments typically operate under live-bed conditions, evaluating their performance under rigid-bed conditions remains relevant in practical applications. For instance, some installations may occur on coarser and more stable beds with limited sediment transport, which more closely resemble rigid-bed behavior.

\section{Conclusion}
\label{sec:5}

We investigated the two-way interactions between a utility-scale VAT and sediment dynamics using a typical riverine sand size. Various TSRs were tested to analyze how turbine-induced wake flow affects bed deformation and how the evolving bed, in turn, influences turbine performance. The ASM was implemented to model the turbine blades, and the immersed boundary method was used to capture the evolving bed geometry. Sediment transport was captured by solving a sediment mass balance equation and using a dual time-stepping scheme and a sand slide model to maintain realistic bed slopes. To that end, comparative simulations under rigid- and live-bed conditions were carried out with the VFS-Geophysics model, which couples hydrodynamics and bed morphodynamics. The simulation results were carefully analyzed to gain insights into the interactions between turbine wake flows and sediment transport in terms of bed evolution as well as turbine wake structure and performance.


Analysis of simulation results under rigid-bed conditions demonstrated that increasing TSR strengthens blade–flow interactions, producing more intense shear layers, elevated turbulence, and sustained wake structures extending farther downstream. Higher TSR confined turbulence to the near-wake and rotor region, raised bed shear stress, and generated a downward-shifting wake core closer to the bed, enhancing erosion potential. Turbulence in the near wake intensified along the blade edges and expanded toward the channel center. The counterclockwise rotation of the blades also introduced an asymmetry in wake flow, skewing turbulence toward the upper sidewall.

Analysis of coupled hydro–morphodynamic simulations under live-bed conditions showed that bed deformation slightly reduced near-bed momentum but had a limited influence beyond approximately $0.3D$ above the bed. Crucially, live-bed effects shortened the wake and accelerated velocity recovery compared to rigid-bed cases. This enhanced recovery was driven by scour beneath the turbine and downstream deposition, which generated near-bed jet flows injecting momentum into the wake core.
Additionally, under live-bed conditions, turbulence becomes concentrated in the near field and around the centerline, particularly as TSR increases. Conversely, TKE decreases near the sidewalls, a mechanism that could help stabilize riverbanks in practical deployments.

The bed-morphodynamics analysis began with distinguishing the dynamic equilibrium state, characterized by the growth of the maximum scour depth along the live bed. At equilibrium, higher TSRs intensified turbulence, deepening scour holes from $-0.0979D$ to $-0.1726D$ and raising sand-bar heights from $0.122D$ to $0.240D$ as TSR increased from $1.6$ to $2.4$, respectively. Increasing the TSR also transported more sediment farther downstream and altered the downstream flow structure. Consequently, in real-world deployments and VAT farms, these changes have important implications for array design and performance, as prior studies have shown that wake–wake interactions within turbine arrays can be substantially detrimental \cite{[148], [149]}. Moreover, bed shear stress analysis confirmed that rotating blades generated strong downstream shear layers, with elevated zones beneath the turbine and near the sidewalls. Large variations in the Shields parameter correlate with major bed deformations, as sediment entrained in high-shear zones is transported and deposited downstream of the turbine, where bed shear stress decreases.

Finally, turbine performance analysis revealed that under rigid-bed conditions, efficiency decreased with TSR reduction, yielding $C_p$ values of $0.329$, $0.320$, and $0.303$ for TSR $=2.4$, $2.0$, and $1.6$, respectively. Under live-bed conditions, $C_p$ decreased across all cases relative to rigid-bed conditions, with efficiency losses of $1.1\%$, $2.2\%$, and $2.5\%$ for TSR $= 1.6$, $2.0$, and $2.4$, respectively. Additionally, analysis of turbine efficiency under rigid-bed conditions suggested an optimum TSR near $2.4$. Under live-bed conditions, however, the optimum TSR shifts to values greater than $2.4$, indicating that sediment–turbine interactions both reduce efficiency and change the TSR associated with peak performance.

\section*{Acknowledgements}
\label{sec:acknowledge}
\noindent This work was supported by grants from the U.S. Department of Energy (DOE)’s Office of Energy Efficiency and Renewable Energy (EERE) under the Water Power Technologies Office (WPTO) Award Numbers DE-EE0009450 and DE-EE00011379. Partial support was provided by NSF (grant number 2233986). The computational resources for the simulations of this study were partially provided by the Institute for Advanced Computational Science at Stony Brook University. The views expressed herein do not necessarily represent the view of the U.S. DOE or the United States Government.

\section*{Author contributions}

\noindent
\textbf{Mehrshad Gholami Anjiraki:} Conceptualization (equal); Data curation (equal); Formal analysis (equal); Investigation (equal); Methodology (equal); Visualization (equal); Writing – original draft (equal); Writing – review \& editing (equal).
\textbf{Mustafa Meriç Aksen:} Investigation (equal); Visualization (equal); Writing – review \& editing (equal). \textbf{Samin Shapourmiandouab:} Investigation (equal); Writing – original draft (equal); Writing – review \& editing (equal). \textbf{Jonathan Craig:} Investigation (equal); 
Writing – original draft (equal); Writing – review \& editing (equal). \textbf{Ali Khosronejad:} Conceptualization (equal); Data curation (equal); Formal analysis (equal); Funding acquisition (lead); Investigation (equal); Methodology (equal); Project administration (lead); Resources (lead); Software (lead); Supervision (lead); Validation (equal); Visualization (equal); Writing – original draft (equal); Writing – review \& editing (equal).

\section*{Data availability statement}
\label{sec:7}
\noindent The software code (VFS-$3.1$ model) (\href{https://doi.org/10.5281/zenodo.15002824}{10.5281/zenodo.15002824}), 
along with the hydrodynamic results (\href{https://doi.org/10.5281/zenodo.15002375}{10.5281/zenodo.15002375}), 
power production data (\href{https://doi.org/10.5281/zenodo.15001388}{10.5281/zenodo.15001388}), 
wake recovery (\href{https://doi.org/10.5281/zenodo.15001454}{10.5281/zenodo.15001454}), 
the instantaneous morphodynamic results (\href{https://doi.org/10.5281/zenodo.15001934}{10.5281/zenodo.15001934}) for the test cases, and the channel and VAT surface files (\href{https://doi.org/10.5281/zenodo.15002280}{10.5281/zenodo.15002280}), 
are available in the Zenodo online repository.

\bibliographystyle{elsarticle-num-names} 
\bibliography{main.bib}

\end{document}